\documentclass[journal]{IEEEtran}
\usepackage{amsmath,amsfonts,amsthm,amssymb,bbm}
\usepackage{algorithm}
\usepackage{algpseudocode}
\usepackage{array}
\usepackage[caption=false,font=tiny,labelfont=sf,textfont=sf]{subfig}
\usepackage{textcomp}
\usepackage{stfloats}
\usepackage{url}
\usepackage{verbatim}
\usepackage{graphicx}
\usepackage{bm}
\usepackage{subfig}
\usepackage{ulem}
\usepackage{color}
\usepackage{diagbox}
\algdef{SE}[DOWHILE]{Do}{doWhile}{\algorithmicdo}[1]{\algorithmicwhile\ #1}%

\newcommand{\bd}{\boldsymbol}
\newcommand{\mb}{\mathbf}
\renewcommand{\vec}[1]{\ensuremath{\boldsymbol{#1}}}
\newcommand{\abs}[1]{\left| #1 \right|}
\newcommand{\norm}[1]{\left\|{#1}\right\|}

\newcommand{\T}{\mathrm{T}}
\renewcommand{\H}{\mathrm{H}}

\newcommand{\cov}{\mathrm{Cov}}
\renewcommand{\j}{\mathbf{j}}

\DeclareMathOperator{\argmax}{\mathrm{argmax}}

\begin{document}
\title{\huge Quasi-Newton OMP Approach for Super-Resolution Channel  Estimation and Extrapolation}

\author{Yi Zeng, Mingguang Han, Xiaoguang Li, and Tiejun Li
\thanks{Y. Zeng, M. Han, and T. Li are with Laboratory of Mathematics and Applied Mathematics, School of Mathematical Sciences,  Center for Machine Learning Research, Peking University, Beijing 100871, P.R. China}
\thanks{Email:  zengyi0427@pku.edu.cn (Y. Zeng), mingguanghan@stu.pku.edu.cn (M. Han),  tieli@pku.edu.cn (T. Li)}
\thanks{X. Li is with MOE-LCSM, School of Mathematics and Statistics, Hunan Normal University, Changsha, Hunan 410081, P. R. China}
\thanks{Email: lixiaoguang@hunnu.edu.cn (X. Li)}
\thanks{Corresponding author: Xiaoguang Li, Tiejun Li}
}

\maketitle

\begin{abstract}
Channel estimation and extrapolation are fundamental issues in MIMO communication systems. In this paper, we proposed the quasi-Newton orthogonal matching pursuit (QNOMP) approach to overcome these issues with high efficiency while maintaining accuracy. The algorithm consists of two stages on the super-resolution recovery: we first performed a cheap on-grid OMP estimation of channel parameters in the sparsity domain (e.g., delay or angle), then an off-grid optimization to achieve the super-resolution. In the off-grid stage, we employed the BFGS  quasi-Newton method to jointly estimate the parameters through a multipath model, which improved the speed and accuracy significantly. Furthermore, we derived the optimal extrapolated solution in the linear minimum mean squared estimator criterion, revealed its connection with Slepian basis, and presented a practical algorithm to realize the extrapolation based on the QNOMP results. Special treatment utilizing the block sparsity nature of the considered channels was also proposed. Numerical experiments on the simulated models and CDL-C channels demonstrated the high performance and low computational complexity of QNOMP.

\end{abstract}

\IEEEpeerreviewmaketitle
\section{Introduction }
\IEEEPARstart{M}{assive} multiple-input multiple-output (MIMO) system is considered to be the most promising technology for wireless communications \cite{MIMO1,MIMO2,MIMO3,MIMO4,MIMO5}. To embrace the potential gains of  MIMO, the accurate estimation of channel state information (CSI) between the transmitter/user and receiver/base station (BS) is a fundamental issue.  In a time division duplex (TDD) system, the users transit the uplink pilots with a prescribed pattern of partial bands in different time slots instead of the whole frequency bands for the communication. The BS side needs to estimate the CSI on the pilots band (i.e., the CSI estimation) and predict the CSI out of pilots bands (i.e., the CSI extrapolation) to identify the transmitted signal. In general, the CSI extrapolation is more challenging than CSI estimation.

It has been revealed \cite{sparse} that although being measured in frequency and space (i.e., the antenna) domain, the CSI has a sparse representation in  the virtual (i.e., delay and angle) domain, which provides a tractable  linear multipath characterization. Utilizing such sparsity provides a more efficient CSI estimation. Based on the estimated virtual parameters, one can easily extrapolates CSI to bands outside of pilots by utilizing the multipath channel model.

Compressed sensing (CS) is an effective technique to reconstruct the sparse solution from a small number of linear measurements.  One popular approach for CS is the greedy type methods, like the orthogonal matching pursuit (OMP) and its different variants \cite{OMP, SP, CoSaMP, ROMP, gOMP, StOMP, SAMP}, for its low computational complexity. Another type of CS methods are based on the convex $\ell_1$ relaxation of $\ell_0$ minimization \cite{CS}. Typical convex relaxation algorithms include the fast iterative shrinkage-thresholding algorithm (FISTA) \cite{FISTA}, alternating direction method of multipliers (ADMM) \cite{ADMM}, iterative re-weighted least squares (IRLS) \cite{IRLS}, etc. Bayesian framework is also popular for sparse reconstruction by incorporating rich sparsity priors. The $\ell_1$ relaxation methods can be formulated in Bayesian framework through the Laplace prior \cite{BCS}. Other CS methods in Bayesian framework include the sparse Bayesian learning (SBL) \cite{SBL}, approximate message passing (AMP) \cite{AMP} and its different extensions \cite{OAMP,VAMP, TCS}.

While the algorithms above  are carried out on a pre-defined grid in virtual domain, they suffer from the off-grid or energy leakage issue. It has been shown  that no matter how fine the grid in the virtual space is, the off-grid mismatch still exists \cite{mismatch}. The off-grid effect degrades the algorithm performance, especially for CSI extrapolation, which results in the super-resolution  study of virtual parameters estimation \cite{SR}.

Among the super-resolution algorithms, the multiple signal classification (MUSIC) \cite{MUSIC} and estimating signal parameters via rotational invariance techniques (ESPRIT) method \cite{ESPRIT} are effective for high signal-to-noise ratio (SNR) case. Atomic norm soft thresholding (AST) \cite{AND} solves a convex optimization on continuous parameter space, which is a continuous limit of LASSO. The Newtonized OMP (NOMP) \cite{NOMP, 2dNOMP} is an OMP type algorithm using the Newton optimization refinement to resolve the basis mismatch issue, which achieves low computational cost and high accuracy.  The off-grid type SBL is also proposed by using a linear approximation of off-grid term or dynamic off-grid model in the SBL framework \cite{OGSBL, RSBL}.

Moreover, the MIMO channel has  specific intrinsic structures \cite{Tse2005}, which can be utilized to further improve the performance of CSI estimation and extrapolation. In particular, the burst sparsity in angular domain is adopted to design the burst LASSO algorithm \cite{Burst}. Similar idea is also used in  structured Turbo-CS algorithm \cite{STCS}. A non-uniform burst-sparsity prior is also introduced in \cite{NonUniform} to handle the off-grid mismatch issue.

In this paper, we propose a two-stage super-resolution algorithm named QNOMP. It can be regarded as an extension of NOMP, which consists of an on-grid sparsity selection stage and an off-grid joint optimization stage. Specially, we apply the Broyden-Fletcher-Goldfarb-Shanno (BFGS)  quasi-Newton optimization in the off-grid stage, which achieves rapid convergence and high accuracy with low computational cost. These QNOMP results are further utilized in the optimal extrapolation of CSI in a seamless manner.

Our major contributions are as follows:
\begin{itemize}
\item\textit{Efficient quasi-Newton optimization in off-grid stage with high efficiency and accuracy}. We resolve the grid-mismatch issue through the off-grid optimization. In this stage, we apply the BFGS type quasi-Newton method, which is matrix inverse free with super-linear convergence rate. This high efficiency allows the joint optimization of all the channel parameters instead of the optimization of single path in each step of NOMP, which significantly improves the estimation speed and accuracy.

\item \textit{Linear optimal extrapolation}. We derive the linear optimal extrapolation for multipath model with the linear minimum mean squared estimator (LMMSE) criterion. Using Gaussian approximation, we can compute this extrapolation efficiently by taking advantage of QNOMP optimization results. This technique further improves the extrapolation performance.

\item \textit{Block sparsity structure exploitation}. We propose a block reweighting technique, which can identify the block sparsity structure of TDD MIMO  channel in the angular domain and achieve more accurate channel estimation and extrapolation. Numerical experiments on the simulated models and CDL-C channels validate the high performance and low computational complexity of QNOMP.
\end{itemize}

The rest of this paper is organized as follows. In Section \uppercase \expandafter {\romannumeral 2}, we introduce the setup of  TDD MIMO  system and the multipath channel model. In Section \uppercase \expandafter {\romannumeral 3}, we introduce our super-resolution algorithm QNOMP in detail. In Section \uppercase \expandafter {\romannumeral 4}, we formulate the extrapolation problem in Bayesian framework and derive the linear optimal extrapolation and its low rank approximation. Moreover, a block reweighting technique for  CSI with block sparsity in the angular domain is also proposed. Numerical experiments and related discussions are presented in Section \uppercase \expandafter {\romannumeral 5}. Finally we make the conclusion.

\textit{Notations:} We denote the matrices and vectors by bold uppercase and lowercase letters, respectively. Scalars are in normal font, and  the italic $i, j, k$ are utilized as subscripts. The superscripts $(\cdot)^\H$ and $(\cdot)^\T$ denote conjugate transpose and transpose, respectively. $|\cdot|$ means the absolute value or modulus of a real/complex number. $\Re(z)$ is the real part of $z$. The Kronecker product is denoted by $\otimes$. $\text{diag}(\bm{x})$ means a diagonal matrix with vector $\bm{x}$ on its diagonal. $CN(\mu,\Sigma)$ denotes complex Gaussian distribution with mean $\mu$ and covariance $\Sigma$. For a given delay $\tau$ or angle $\theta$,  we denote its corresponding frequency atom  as $\vec{a}_\tau(\tau):= [1, e^{-\j 2\pi\tau\Delta f}, \cdots, e^{-\j 2\pi(M-1)\tau\Delta f}]^\T$ and $\vec{a}_\theta(\theta):= [1, e^{-\j 2\pi\theta}, \cdots, e^{-\j 2\pi(M-1)\theta}]^\T$, where $\j^2=-1$ is the imaginary unit.  Given a set of delay (or angle) $\Omega=\{\tau_1, \tau_2,\dots, \tau_K\}$ (or $\Theta=\{\theta_1, \theta_2,\dots, \theta_L\}$), we always assume they are sorted in a proper order and write them in a vector $\vec{\tau}=[\tau_1, \tau_2,\dots, \tau_K]^\T$ (or $\vec{\theta}=[\theta_1, \theta_2,\dots, \theta_L]^\T$). The dictionary matrix corresponding to a pair of delay and angle vector $(\vec{\tau}, \vec{\theta})$ is denoted as $A(\vec{\tau},\vec{\theta})=[\vec{a}_{\tau}(\tau_i)\otimes\vec{a}_{\theta}(\theta_j)]$ for $i=1:K; j=1:L$, where $1:K=\{1,2,\ldots,K\}$ and same rule applies to similar notations.

\section{The MIMO Channel Model}

We consider a frequency-selective  TDD MIMO  channel. Suppose the BS is equipped with a uniform linear array (ULA) with $N$ antennas. The user sends $M$ pilots  $x_i \in\mathbb{C}$, $i=1:M$ on $M$ subcarriers with subcarrier frequency spacing $\Delta f$. The BS receives signal $\mb y_k \in \mathbb{C}^{N}$ as
    \begin{equation}\label{eq:model}
      \mb y_k=\mb h_kx_k+\mb w_k, \quad k=1:M,
    \end{equation}
where $\mb h_k$ denotes the CSI in space domain and $\{\mb w_k\}_k$ are independent additive Gaussian white noise (AGWN). The CSI $\mb h_k$ can be expressed through a multipath model
\begin{equation}\label{eq:multipath1}
\mb h_k=\sum_{i=1}^{L} \beta_i e^{-\j 2\pi \tau_i k\Delta f} \mb{a}_\theta\left(\theta_{i}\right),\quad k=1:M,
\end{equation}
where $L$ is the number of signal paths, $\beta_i$ the complex gain,  $\tau_i$  the time delay, and $\theta_{i}\propto\cos \phi_i$ the directional cosine of the angle-of-arrival (AoA) $\phi_i$ of the $i$-th path, respectively. More precisely, $\mb a_{\theta}(\theta)$ is the steering vector
\begin{equation}\label{eq:steeringvector}
\mb a_\theta(\theta)=\left[1, e^{-\j 2 \pi \theta}, \cdots, e^{-\j 2 \pi\left(N-1\right) \theta}\right]^\T,
\end{equation}
where $\theta=d/\lambda\cdot \cos \phi$, $d$ is the antenna spacing, $\lambda$ is the subcarrier wavelength,  and $\phi$ is the AoA. We simply take  $d=\lambda/2$ in this paper, then  $\theta\in [-1/2,1/2]$ and we take $\phi\in[0,\pi]$. Define the frequency response vector
\[\mb a_\tau (\tau)=\left[1, e^{-\j 2 \pi \tau \Delta f}, \dots, e^{-\j  2 \pi\tau\left(M-1\right)\Delta f }\right]^\T.\]
 The frequency-antenna impulse response can be expressed by combining the steering vector and frequency response vector 
\[\mb a(\tau_i,\theta_i)=\mb a_\tau(\tau_i)\otimes \mb a_\theta(\theta_i)\in \mathbb{C}^{NM}.\]
Denote $\bd \beta=[\beta_1,\dots,\beta_{L}]\in \mathbb{C}^{L}$. The channel response $\mb h=[\mb h_1^\T, \mb h_2^\T, \dots, \mb h_M^\T]^\T\in\mathbb{C}^{NM}$ can be expressed in a matrix form:
\begin{equation}\label{eq: multipath}
\mb h=\sum_{i=1}^{L}\beta_i \mb a(\tau_i,\theta_i)=\mb A(\bd \tau,\bd \theta)\bd \beta,
\end{equation}
    where $\bd \tau=[\tau_1,\cdots,\tau_{L}]^\T$, $\bd \theta=[\theta_1,\cdots,\theta_{L}]^\T$ in $\mathbb{R}^{L}$. The sensing matrix $\mb A(\bd \tau,\bd \theta)=[\mb a(\tau_1,\theta_1),\cdots,\mb a(\tau_{L},\theta_{L})]\in \mathbb{C}^{NM\times L}$.

Let $\mb x = [x_1, x_2, \dots, x_M]\in\mathbb{C}^M$ denote the collection of pilots. We can rewrite the received signal $\mb y=[\mb y_1^\T,\dots,\mb y_m^\T]^\T\in\mathbb{C}^{NM}$ into a matrix form
\begin{equation}\label{eq:ymat}
\mb y=\operatorname{diag}(\mb x\otimes 1_N)\mb h+\mb w,
\end{equation}
where $\mb w \sim CN(0,\sigma^2 I_{NM})$ is AGWN.

Assuming $\abs{x_i}=1$, the least squares estimation of $\mb h$ based on the observation $\mb y$ is
\begin{align}\label{eq: ls}
\mb h'&=\operatorname{diag}(\mb x\otimes 1_N)^{-1}\mb y=\mb h+\operatorname{diag}(\mb x\otimes 1_N)^{-1}\mb w \\
&=\mb h + \mb w',\mb w'\sim CN(0,\sigma^2  I_{NM}). \nonumber
\end{align}
The super-resolution recovery of channel state is to estimate the parameters $(\vec{\beta}, \vec{\tau}, \vec{\theta})$. With this estimation, one can extrapolate the channel states to the unobserved frequency bands  by model \eqref{eq:multipath1}.

\section{Super-Resolution Algorithm: QNOMP}\label{sec:main}

Taking advantage of the sparsity nature of the channel in multipath model, the channel parameters can be estimated by solving the following sparse optimization problem with loss
\begin{equation}\label{eq:prob}
\min_{\bd \tau,\bd \theta,\bd \beta} \mathcal{L}=\frac{1}{\sigma^2}\|\mb h'-\mb A(\bd \tau,\bd \theta)\bd \beta\|_2^2, \quad \text{s.t. $\bd \beta$ is sparse}
\end{equation}
based on the $\mb h'$ defined in \eqref{eq: ls}. We propose a two-stage algorithm for this problem. In the first stage (nonstandard \textit{on-grid stage}), we determine the sparsity and give a rough estimate of parameters by an improved OMP-type method with a \textit{multi-resolution} on-grid sparsity selection and a slight off-grid nonlinear refinement. After the first stage, an \textit{off-grid stage} is followed to obtain more accurate parameters by the BFGS quasi-Newton method. We call this approach quasi-Newton OMP (QNOMP) algorithm.

OMP is an efficient on-grid method for sparsity selection and channel estimation. Given $M$ pilots and corresponding observation $\mb y$, we introduce a standard grid on delay and angular domain
\begin{equation}\label{eq:grid0}
\begin{aligned}
\vec{\tau}^0&=\Delta^\tau_{\text{dft}}\cdot[0,1,\dots, M-1],\\
\vec{\theta}^0&=\Delta^\theta_{\text{dft}}\cdot[-N/2,\dots, N/2-1],
\end{aligned}
\end{equation}
respectively, where the grid size $\Delta^\tau_{\text{dft}}=1/(M\Delta f)$, $\Delta^\theta_{\text{dft}}=1/N$. Define the sensing matrix $\mb A(\vec{\tau}^0, \vec{\theta}^0)=[\mb a(\tau^0_i)\otimes \mb a(\theta^0_j)]\in\mathbb{C}^{NM\times NM}$ for $i=0:M, j=-N/2:(N/2-1)$. The OMP algorithm is to consider the on-grid version of 
\eqref{eq:prob}:
\begin{equation}
\min_{\bd \beta} \mathcal{L}^{\text{on}}=\frac{1}{\sigma^2}\|\mb h'-\mb A(\bd \tau^0,\bd \theta^0)\bd \beta\|_2^2 \quad \text{s.t. $\bd \beta$ is sparse}.
\end{equation}
One first initializes the residual vector $\vec{r}=\mb h'$ and empty parameter sets $\hat{\vec{\tau}}=\hat{\vec{\theta}}=\emptyset$, then in each OMP iteration, one estimates the delay and angle of a certain path, which will be added to the corresponding parameter set. To be specific, one finds the column of $\mb A(\bd \tau^0,\bd \theta^0)$ that most closely correlated to $\vec{r}$, i.e., finding the index $(i,j)=\argmax_{i,j}\abs{\vec{r}^\H\mb a_\tau(\tau^0_i)\otimes \mb a_\theta(\theta_j^0)}$. The sets $\hat{\vec{\tau}}$, $\hat{\vec{\theta}}$ are then updated as $\hat{\vec{\tau}}\leftarrow \hat{\vec{\tau}}\cup\{\tau^0_i\}$, $\hat{\vec{\theta}}\leftarrow \hat{\vec{\theta}}\cup\{\theta^0_j\}$. Based on the estimated $\hat{\vec{\tau}}$ and $\hat{\vec{\theta}}$, the gain $\vec{\beta}$ is determined by a least squares estimation
    \begin{equation}\label{eq:beta}
      \hat{\vec{\beta}}=\mb A(\hat{\vec{\tau}}, \hat{\vec{\theta}})^\dagger\vec{r}.
    \end{equation}
The residual vector is updated as
    \begin{equation}\label{eq:residual}
      \vec{r}\leftarrow\mb h'-\mb A(\hat{\vec{\tau}}, \hat{\vec{\theta}})\hat{\vec{\beta}}.
    \end{equation}
OMP runs the above procedure iteratively until some stopping criterion is satisfied. The number of iterations $N_p$ is the estimated number of propagation paths.

OMP is a greedy method. In each iteration, it finds the optimal delay and angle for a single path. However, these sequential optimization does not necessarily achieve joint optimum for multipaths. Meanwhile, the resolution of OMP algorithm is restricted by the grid size $\Delta_{\text{dft}}^\tau$, $\Delta_{\text{dft}}^\theta$, and the true location of $\tau_i, \theta_i$ of a certain path may not lie on the pre-defined grid $\vec{\tau}^0, \vec{\theta}^0$. This grid mismatch significantly degrades the performance of super-resolution and extrapolation. To overcome this issue, we introduce a two-stage QNOMP strategy consist of an  on-grid multi-resolution optimization stage with local refinement, and an off-grid BFGS joint optimization stage, both of which are proceeded in an interweaving  manner.

\subsection{On-Grid Multi-resolution OMP for Sparsity Selection}\label{on grid}

To obtain a better estimation of delay and angle, we first apply a locally refined multi-resolution technique in each OMP iteration. The OMP estimation $\hat{\tau}^0={\tau}^0_i, \hat{\theta}^0={\theta}^0_j$ serves as the initial guess of this procedure. Given the estimation $\hat{\tau}^n, \hat{\theta}^n$ after the $n$-th refinement iteration, we construct a locally refined grid
    \begin{equation}\label{eq:gridn}
      \begin{aligned}
      \tilde{\vec{\tau}}^n &= \hat{\tau}^n+\frac{\Delta^\tau_{\text{dft}}}{k_1^{n+1}}[-k_1,\dots,0,\dots,k_1]\in\mathbb{R}^{2k_1+1}, \\
      \tilde{\vec{\theta}}^n& = \hat{\theta}^n+\frac{\Delta^\theta_{\text{dft}}}{k_2^{n+1}}[-k_2,\dots,0,\dots,k_2]\in\mathbb{R}^{2k_2+1},
      \end{aligned}
    \end{equation}
    where the two integers $k_1, k_2\geq 1$ indicate the refinement rate. The refined estimation is $\hat{\tau}^{n+1}=\tilde{\tau}_i^{n}, \hat{\theta}^{n+1}=\tilde{\theta}_j^{n}$, where the index $(i,j)=\argmax_{i,j} |\vec{r}^\H\mb a_\tau(\tilde{\tau}^n_i)\otimes \mb a_\theta(\tilde{\theta}^n_j)|$. The procedure can be applied iteratively to achieve a high precision, although practically one or two times of refinement can lead to a satisfying result.

    Since the local refinement grid of each iteration has a constant size $J=(2k_1+1)(2k_2+1)$, this local refinement procedure makes the resolution of delay and angle estimations to be $\Delta_{\text{dft}}^\tau/k_1^n$ and  $\Delta_{\text{dft}}^\theta/k_2^n$ by an additional $O(nJ)+O(nJ\min\{M,N\})$ computation, while the resolution of conventional OMP is $\Delta_{\text{dft}}^\tau$ and $\Delta_{\text{dft}}^\theta$. We summarize the local refinement procedure in Algorithm~\ref{alg:1}.
\begin{algorithm}
\caption{On-Grid Multi-resolution OMP Procedure}\label{alg:1}
\begin{algorithmic}[1]
\State \textbf{Input:} Residual vector $\vec{r}$, OMP grid size $\Delta_{\text{dft}}^{\tau}$, $\Delta_{\text{dft}}^{\theta}$ and estimation $\hat{\tau}^0$, $\hat{\theta}^0$, refinement rate $k_1, k_2$, maximum iteration number $n_{\text{max}}$
\State \textbf{Output:} Refined estimation $\hat{\tau}$ and $\hat{\theta}$
\State \textbf{Initialize:} $\hat{\tau}=\hat{\tau}^0$, $\hat{\theta}=\hat{\theta}^0$, $n=1$
\While{$n < n_{\text{max}}$}
    \State Construct grid $\tilde{\vec{\tau}}$ and $\tilde{\vec{\theta}}$ by \eqref{eq:gridn}
    \State Compute $R_{ij}=\vec{r}^\H\mb a_\tau(\tilde{\tau}_i)\otimes \mb a_\theta(\tilde{\theta}_j)$
    \State Find $(i,j)=\argmax_{i,j}\abs{R_{ij}}$ on the refined grids
    \State Update $\hat{\tau} \leftarrow \tilde{\tau}_i$, $\hat{\theta} \leftarrow \tilde{\theta}_j$
    \State $n \leftarrow n+1$
\EndWhile\\
        \Return $\hat{\tau}$, $\hat{\theta}$
\end{algorithmic}
\end{algorithm}

Like conventional OMP algorithm, this on-grid module refines the estimation of delay and angle of a single path, which leaves other estimated path parameters unchanged.

\subsection{Off-Grid BFGS Parameter Optimization}\label{sec:off-grid}

The on-grid stage gives us an estimation of $\tau_i$ and $\theta_j$ on the grid $\tilde{\vec{\tau}}^n$, $\tilde{\vec{\theta}}^n$. However, the true location $\tau_i$, $\theta_i$ may not lie on this grid. To deal with this grid mismatch issue, we treat $\vec{\tau}$, $\vec{\theta}$ in \eqref{eq:prob} as continuous variables and optimize them by BFGS Quasi-Newton algorithm.

In the $k$-th iteration of OMP, we have selected $k$ propagation paths with delay $\hat{\vec{\tau}}\in\mathbb{R}^k$ and angle $\hat{\vec{\theta}}\in\mathbb{R}^k$. Note that given a set of $(\vec{\tau}, \vec{\theta})$, the gain $\vec{\beta}$ can be estimated by least squares estimation $\hat{\vec{\beta}}=\hat{\vec{\beta}}(\vec{\tau}, \vec{\theta})=\mb A^\dagger(\vec{\tau}, \vec{\theta})\vec{h}'$. We are left to solve the continuous optimization problem
\begin{equation}\label{eq:optim}
\min_{\vec{\tau},\vec{\theta}} \mathcal{L}(\vec{\tau},\vec{\theta})=\frac{1}{\sigma^2}\norm{\vec{h}'-\mb A(\vec{\tau}, \vec{\theta})\hat{\vec{\beta}}(\vec{\tau},\vec{\theta})}_2^2.
    \end{equation}

We take the BFGS algorithm, one of the most effective quasi-Newton method, to perform the optimization \cite{Nocedal2018NumericalO}.  It utilizes the historically computed gradient information to approximate the inverse of Hessian matrix which avoids computing the matrix inverse. Applying BFGS to \eqref{eq:optim}, the scheme reads
\begin{equation}\label{eq:BFGS1}
\begin{aligned}
\vec{\tau}^{n+1} = \vec{\tau}^n & + \alpha_n H^\tau_n\vec{g}_n^\tau,\\
\vec{\theta}^{n+1} = \vec{\theta}^n &+ \alpha_n H^\theta_n\vec{g}_n^\theta, \\
H^\tau_{n+1} = H^\tau_{n} & + \left(1 + \frac{[\vec{\delta g}_n^{\tau}]^\T H_n^\tau \vec{\delta g}_n^\tau}{[\vec{\delta g}_n^\tau]^\T\vec{\delta \tau}_n} \right)\cdot\frac{\vec{\delta \tau}_n\vec{\delta \tau}_n^\T}{[\vec{\delta g}_n^\tau]^\T\vec{\delta \tau}_n} \\
&- \frac{\vec{\delta\tau}_n[\vec{\delta g}_n^\tau]^\T H_n^\tau + H_n^\tau\vec{\delta g}_n^\tau\vec{\delta\tau}_n^\T}{[\vec{\delta g}_n^\tau]^\T\vec{\delta \tau}_n},       \\
H^\theta_{n+1} = H^\theta_{n} & + \left(1 + \frac{[\vec{\delta g}_n^\theta]^\T H_n^\theta \vec{\delta g}_n^\theta}{[\vec{\delta g}_n^\theta]^\T\vec{\delta \theta}_n} \right)\cdot\frac{\vec{\delta \theta}_n\vec{\delta \theta}_n^\T}{[\vec{\delta g}_n^\theta]^\T\vec{\delta \theta}_n} \\
&- \frac{\vec{\delta\theta}_n[\vec{\delta g}_n^\theta]^\T H_n^\theta + H_n^\theta\vec{\delta g}_n^\theta\vec{\delta\theta}_n^\T}{[\vec{\delta g}_n^\theta]^\T\vec{\delta \theta}_n}
\end{aligned}
\end{equation}
with  the definition  $\vec{\delta\tau}_n = \vec{\tau}^{n+1} - \vec{\tau}^n$, $\vec{\delta\theta}_n = \vec{\theta}^{n+1} - \vec{\theta}^n$, and $\vec{\delta g}_n^{\tau,\theta} = \vec{g}_{n+1}^{\tau,\theta} -  \vec{g}_{n}^{\tau,\theta}$.

In \eqref{eq:BFGS1}, the vectors $\vec{g}_n^\tau$, $\vec{g}_n^\theta$ indicate the gradient of $\vec{\tau}$ and $\vec{\theta}$ at the $n$-th iteration step $(\vec{\tau}^n, \vec{\theta}^n)$, respectively. The step size $\alpha_n>0$ can be determined by the Armijo line search rule \cite{Nocedal2018NumericalO}. The matrix $H_n^\tau, H_n^\theta\in\mathbb{R}^{k\times k}$ is the approximated inverse Hessian matrix. One can derive $\nabla_{\tau} \mb a_\tau(\tau)$, $\nabla_{\theta} \mb a_\theta(\theta)$ and related variables as below
\begin{equation}\label{eq:gradient1}
\begin{aligned}
&\nabla_{\bd\tau} \mb A=\nabla_{\tau} \mb a_\tau\otimes \mb a_\theta ,\ \nabla_{\bd\theta} \mb A  = \mb a_\tau\otimes \nabla_{\theta} \mb a_\theta, \\
&\vec{g}_n^\tau = \nabla_{\bd \tau} \mathcal{L} = \frac{2}{\sigma^2}\Re\big[\bd \hat{\vec{\beta}}^*\odot \nabla_{\bd \tau} \mb A^\H \mb (\mb A\hat{\vec{\beta}} - \vec{h}')\big],\\
&\vec{g}_n^\theta  = \nabla_{\bd \theta} \mathcal{L}= \frac{2}{\sigma^2}\Re\big[\bd \hat{\vec{\beta}}^*\odot \nabla_{\bd \theta} \mb A^\H \mb (\mb A\hat{\vec{\beta}} - \vec{h}')\big],
\end{aligned}
\end{equation}
where the above equations are evaluated at $(\vec{\tau},\vec{\theta})=(\vec{\tau}^n,\vec{\theta}^n)$, and $\hat{\vec{\beta}}^*$ is the conjugate of complex vector $\hat{\vec{\beta}}$. To start the iteration, the OMP on-grid estimation $(\hat{\vec{\tau}}, \hat{\vec{\theta}})$ can serve as the initial value for $(\vec{\tau},\vec{\theta})$, and the initial value of matrix $H$ can be chosen as the identity matrix $I_k$. The off-grid  BFGS refinement algorithm is shown in Algorithm~\ref{alg:2}.

    \begin{algorithm}
\caption{Off-Grid BFGS Refinement}\label{alg:2}
\begin{algorithmic}[1]
\State \textbf{Input:} CSI $\vec{h}'$, OMP estimation $\hat{\vec{\tau}}, \hat{\vec{\theta}}\in\mathbb{R}^k$, maximum iteration $n^{\text{in}}$
\State \textbf{Output:} Refined estimation $\vec{\tau}$, $\vec{\theta}$
\State \textbf{Initialize:} $\vec{\tau}=\hat{\vec{\tau}}$, $\vec{\theta}=\hat{\vec{\theta}}$, , $n=1$
\While{$n < n^{\text{in}}$}
            \State Compute $\hat{\vec{\beta}} = \mb A^\dagger(\vec{\tau}, \vec{\theta})\vec{h}'$
    \State Using \eqref{eq:gradient1} to compute the gradient information
    \State Using Armijo rule to determine step size $\alpha$
    \State Update $\vec{\tau}$ and $\vec{\theta}$ by scheme \eqref{eq:BFGS1}.
    \State $n \leftarrow n+1$
\EndWhile\\
        \Return $\vec{\tau}$, $\vec{\theta}$
\end{algorithmic}
\end{algorithm}

Unlike the on-grid module, in the $k$-th iteration, the off-grid BFGS refinement  optimizes the delays and angles of all the $k$ estimated paths jointly. So this module not only  improves the single path estimation from the on-grid module, but also corrects the entire estimated multipath parameters so one can alleviate the issue of being stuck at the local minimum by sequential optimization.

\subsection{The Stopping Criterion of OMP Iteration}
    
The OMP iteration is executed  until the residual vector $\vec{r}$ reaches the noise level. We adopt the constant false alarm rate (CFAR) stopping criterion \cite{NOMP} to measure whether the residual
meets the noise level. Suppose we have got the accurate value of channel parameters $\vec{\tau}, \vec{\theta}$ and $\vec{\beta}$, the residual of the reconstructed signal satisfies $\mathcal{F}^{-1}(\vec{r})=\mb A(\vec{\tau}^0, \vec{\theta}^0)^\H\mb w'/MN$. Due to the orthogonality of $\mb A$, $\tilde{\vec{w}}=\mb A^\H\mb w/\sqrt{MN}$ is still an AGN with covariance $\sigma^2 I$. It is straightforward to show that
\begin{equation*}
\begin{aligned}
&\mathbb{P}\left\{\norm{\mathcal{F}^{-1}(\vec{r})}_{\infty}^2<T_v\big|\vec{\tau},\vec{\theta},\vec{\beta}\right\} = \mathbb{P}\left\{\|\tilde{\vec{w}}\|_{\infty}^2 < MNT_v \right\} \nonumber\\ &= \mathbb{P}\big\{ \abs{\tilde{w}_1}^2 < MNT_v \big\}^{MN} = \left(1 - e^{-\frac{MN T_v}{\sigma^2}}\right)^{MN},
\end{aligned}
\end{equation*}
where $T_v$ is a constant we are going to determine. Let $T_v = \sigma^2(\log(MN)+Z)/MN$. We have $(1 - e^{-MN T_v/\sigma^2})^{MN}=\left(1 - (MN)^{-1}e^{-Z}\right)^{MN}\approx \exp\left(-\exp(-Z)\right)$ when $MN$ is large. Thus we have
\[\mathbb{P}\left\{\norm{\mathcal{F}^{-1}(\vec{r})}_{\infty}^2>T_v\big|\hat{\vec{h}}=\vec{h}^r \right\}\approx 1-\exp\big(-\exp(-Z)\big).\]
Given a small probability $P_{\text{fa}}$, we can ensure
\begin{equation*}
\mathbb{P}\left\{\norm{\mathcal{F}^{-1}(\vec{y}_r)}_{\infty}^2>T_v\big|\vec{\tau},\vec{\theta},\vec{\beta}\right\}<P_{\text{fa}}
\end{equation*}
by choosing $Z =\log(MN) - \log(-\log(1-P_{\text{fa}}))$, namely
\begin{equation}\label{eq:Tv}
   T_v=(MN)^{-1}\sigma^2(\log(MN) - \log(-\log(1-P_{\text{fa}}))).
\end{equation}
 Therefore, if we found $\norm{\mathcal{F}^{-1}(\vec{r})}^2_{\infty}>MNT_v$, we can reject the original hypothesis that we have found the correct channel parameters, which means the reconstructed signal does not meet the noise level so we should run more OMP iterations.

\subsection{The Joint BFGS Stage}

The modified OMP iteration introduced in Sections \ref{on grid} and \ref{sec:off-grid} determines the number of propagation paths and gives a estimation of channel parameters iteratively. After this stage, we fix the number of paths $N_p$, thus the gains $\vec{\beta}$ will be nonzero. Now the problem becomes a parameter optimization without any sparse constraints. We correct all the delays and angles further with the BFGS algorithm again. In detail, we are going to solve the regularized problem
   \begin{equation}\label{eq: loss}
\min_{\bd \tau,\bd \theta} \mathcal{L}(\bd \tau,\bd \theta)=\frac{1}{\sigma^2}\|\mb h'-\mb A(\bd \tau,\bd \theta)\hat{\bd \beta}\|_2^2+\hat{\bd \beta}^\H \mb E^{-1}\hat{\bd \beta},
\end{equation}
    by BFGS method, where the optimal $\vec{\beta}$ is given by
    \begin{equation}\label{eq:regbeta}
      \hat{\vec{\beta}} = (\mb A^\H\mb A+\sigma^2 \mb E^{-1})^{-1}\mb A^\H \mb h'.
    \end{equation}

In practice, we choose the regularizer $\mb E\in\mathbb{R}^{N_p\times N_p}$ as $\lambda I$ $(\lambda>0)$ for convenience. The regularizer can be also interpreted in a Bayesian framework and improved with this perspective. The choice of $\mb E$ will be further discussed in detail in next section.

The BFGS scheme to solve \eqref{eq: loss} is the same as \eqref{eq:BFGS1}, in which $\hat{\vec{\beta}}$ is set as \eqref{eq:regbeta} and $k = N_p$. Let us summarize the two-stage QNOMP algorithm in Algorithm~\ref{alg:3} for clarity.
    \begin{algorithm}
\caption{QNOMP Algorithm}\label{alg:3}
\begin{algorithmic}[1]
\State \textbf{Input:} CSI $\vec{h}'$, noise level $\sigma$, maximum iteration of local refinement $n^{\text{LR}}$, maximum iteration of BFGS $n^{\text{in}}$ and $n^{\text{out}}$, the refinement rate $k_1, k_2$, false alarm probability $P_{\text{fa}}$
\State \textbf{Output:} Channel parameter estimations $\vec{\tau}$, $\vec{\theta}$ and $\vec{\beta}$
\State \textbf{Initialize:} $\vec{r}=\mb h'$, $\vec{\tau}=\vec{\theta}=\emptyset$, $k=1$, $n=1$
        \Do
            \State Applying Algorithm \ref{alg:1} $n^{\text{LR}}$ steps to get  $(\hat{\vec{\tau}}, \hat{\vec{\theta}})$
            \State Applying Algorithm \ref{alg:2} $n^{\text{in}}$ steps to get  $(\vec{\tau}, \vec{\theta})$
            \State Update $\vec{\beta}=\mb A^\dagger(\vec{\tau}, \vec{\theta})\vec{r}$ and $\vec{r} = \mb h' - \mb A(\vec{\tau}, \vec{\theta})\vec{\beta}$
            \State $k\leftarrow k+1$
\doWhile{$\|\mathcal{F}^{-1}(\vec{r})\|_{\infty}^2>MN T_v$} as in \eqref{eq:Tv}
        \State Set $N_p=k$ and the regularizer $\mb E$
        \While{$n<n^{\text{out}}$}
            \State Using \eqref{eq:BFGS1}, \eqref{eq:gradient1} with $\hat{\vec{\beta}}$ in \eqref{eq:regbeta} to get  $\vec{\tau}$ and $\vec{\theta}$
            \State $n\leftarrow n+1$
        \EndWhile
        \State Let $\vec{\beta} = (\mb A^\H(\vec{\tau},\vec{\theta})\mb A(\vec{\tau},\vec{\theta})+\sigma^2 \mb E^{-1})^{-1}\mb A^\H(\vec{\tau},\vec{\theta}) \mb h'$.\\
        \Return $(\vec{\tau}, \vec{\theta}, \vec{\beta})$
\end{algorithmic}
\end{algorithm}

\subsection{Computational Cost and Remarks}

In this section we analyze the computational cost of QNOMP. In Algorithm~\ref{alg:1}, the cost of Step 6 is $O(NM)$ for each $(i,j)$ pair. Since $\mb a_\tau\otimes \mb a_\theta$ has a tensor structure, the total cost of all $R_{ij}$ can be reduced to $O(NM\min\{M,N\})$ for $k_1=k_2=1$. In Step 7, the cost of finding the maximum of $\abs{R_{ij}}$ is $J=(2k_1+1)(2k_2+1)$. So the cost of each refinement step is $O(NM\min\{M,N\} + J)$. The total cost of Algorithm~\ref{alg:1} is $O((NM+n^{\text{LR}}J)\min\{M,N\})$, where $n^{\text{LR}}$ is the number of refinement steps. The local refinement procedure is very efficient. To achieve the same resolution, the conventional OMP method requires $O(k_1^nk_2^nNM\min\{M,N\})$ computation.

Since almost all optimization algorithms need the gradient information, we only provide the additional computational cost brought by BFGS algorithm. Compared with gradient based methods such as gradient descent, BFGS scheme only needs to compute the update of matrix $H$. The update rule of $H$ is composed by a series of matrix-vector product. In the $k$-th OMP iteration, $H\in\mathbb{R}^{k\times k}$. the cost of updating $H$ is $O(k^2)$ in each step.  Thus the total additional cost brought by BFGS is $n^{\text{in}}\cdot\sum_{k=1}^{N_p}O(k^2)=n^{\text{in}}\cdot O(N_p^3)$, where $n^{\text{in}}$ is number of BFGS steps. Since BFGS is a super-linear algorithm which converges very fast, the BFGS steps $n^{\text{in}}$ can be very small. Moreover, because the channel is sparse, $N_p$ is also small. The additional cost is almost a minor constant. Thus, employing BFGS is highly efficient in this scenario. It introduces little additional cost and achieves faster convergence. In practice, the solutions obtained by BFGS often have higher precision.

We have analyzed the cost of Steps 5 and 6 in QNOMP. The cost of Step 12 is similar as Step 6. Step 7 is the computation of pseudo-inverse with cost $O(k^3+k^2(M+N)+kNM)$ for the $k$-th OMP iteration. So the total cost is $\sum_{k=1}^{N_p}O(k^3+k^2(M+N)+kNM)=O(N_p^4+N_p^3(M+N)+N_p^2NM)$. Meanwhile, Step 15 introduces an additional cost $O(N_p^3+N_p^2(M+N)+N_pNM)$. Since $N_p$ is small, the total cost of these steps is roughly $O(N_p^2NM)$.

Overall, we conclude that QNOMP is efficient. Compared with conventional OMP method, it only requires  almost constant additional cost but can achieve a much more accurate estimation of channel parameters and a powerful extrapolation ability, which will be shown in later numerical results.
    
We have some final remarks on QNOMP:
\begin{enumerate}
\item QNOMP is an improvement and extension of the NOMP method \cite{NOMP,2dNOMP}. In each OMP iteration, we add a multi-resolution refinement module to enhance the performance. The key improvement of QNOMP is that we use BFGS instead of Newton method to compute the off-grid estimation. BFGS allows us to optimize all the estimated path parameters jointly, while in NOMP, it is difficult to do this  since it has to compute the inverse of Hessian matrix in each Newton step. The joint optimization also alleviates the sequentially local minimum issue. Moreover, employing a full-variable BFGS optimization after OMP stage can yield more accurate estimates in practice.
\item In conventional OMP method, the residual vector $\vec{r}$ is orthogonal to the selected column of sensing matrix. Each column of sensing matrix can be selected by OMP at most once. However, due to the off-grid optimization, the orthogonality is lost in QNOMP. So some $(\tau_i^0, \theta_j^0)$ may be selected by on-grid OMP step for multiple times. This feature brings benefit for estimation. Fixing grid size $\Delta_{\text{dft}}^\tau, \Delta_{\text{dft}}^\theta$, it is difficult to distinguish two paths which are very close. The multiresolution refinement module and off-grid BFGS optimization make it possible to select paths whose distance smaller than $\Delta_{\text{dft}}^\tau, \Delta_{\text{dft}}^\theta$ in successive steps.
\item We do not apply the regularization $\mb E$ in the OMP stage to ensure the convergence of OMP iterations. As in conventional OMP, the norm of residual vector $\mb r$ is monotonically decreasing as long as we use a monotonic line search without regularization. In BFGS stage after OMP, the parameters and the number $N_p$ is fixed, so we can add the regularization to perform further optimization.
\item The number of BFGS steps in OMP iteration need not to be large. Practically, two or three steps is enough to give a satisfying result. In BFGS stage after OMP iteration, a few more BFGS steps are required to find a more accurate channel estimation. Moreover, one can select several indices with the largest correlations in OMP iteration. Practice shows that selecting the index with the largest correlation for local refinement is already enough to obtain a high-resolution result.
\end{enumerate}

\section{Linear Optimal Extrapolation and Improvement of QNOMP}

With the optimized parameters $(\bd \tau,\bd \theta,\bd \beta)$, one can get the extrapolation on different frequency bands by substituting them into channel model \eqref{eq:multipath1}. To measure the accuracy of extrapolation, one should consider the approximation error on the whole frequency domain rather than given pilots. Since we will only consider the extrapolation of the channel in frequency domain, we will ignore the spatial domain and focus on the one-dimensional frequency domain channel with the virtual delay domain in Sections \ref{Sec:4A} and \ref{Sec:4B}, then switch back to the case with spatial domain in Section \ref{Sec:4C}.

Let the extrapolated frequency bands be $f_1,\dots, f_K$ and the corresponding response vector $\mb a_e(\tau)=[1, e^{-\mb j 2\pi\tau f_i},\dots, e^{-\j 2\pi\tau(K-1) f_i}]$ for $i=1:K$. For simplicity, we assume $f_i = (M+i-1)\Delta f$. The following results can be extended to general cases without difficulty.

To distinguish the frequency bands for estimation or extrapolation, we denote $\mb A_0=\mb A(\vec{\tau})=[\mb a_\tau(\tau_1),\dots, \mb a_\tau(\tau_{N_p})]\in\mathbb{C}^{M\times N_p}$  and  $\mb A_e=\mb A_e(\vec{\tau})=[\mb a_e(\tau_1), \dots, \mb a_e(\tau_{N_p})]\in \mathbb{C}^{K\times N_p}$ the sensing matrices for estimation and extrapolation, respectively.

The channel parameter estimation can be modeled in Bayesian framework. Given the probability model \eqref{eq: ls}, the posterior distribution of $(\vec{\tau}, \vec{\beta}|\mb h')$ in channel model \eqref{eq:multipath1} can be expressed as
\begin{equation}\label{eq:bayesian}
\begin{aligned}
\log &P(\vec{\tau},\vec{\beta}|\mb h') = \log P(\mb h'|\vec{\tau},\vec{\beta}) + \log P(\vec{\tau},\vec{\beta})\\
&-\log \int P(\vec{\tau},\vec{\beta},\mb h')d\vec{\tau}\vec{\beta}\mb h'\\
&= -\frac{1}{\sigma^2}\norm{\mb h' - \mb A_0\vec{\beta}}_2^2 + \log P(\vec{\tau},\vec{\beta}) + \text{Const.},
\end{aligned}
\end{equation}
where $P(\vec{\tau},\vec{\beta})$ is the prior distribution of $(\vec{\tau},\vec{\beta})$ and the additional constant is irrelevant to the estimation. Assuming $\vec{\tau}$ and $\vec{\beta}$ are independent, $\vec{\tau}$ has a non-informative prior and the prior of $\vec{\beta}$ is $CN(0,\mb E)$, the posterior reads
\begin{equation}\label{eq:posterior}
\log P(\vec{\tau}|\mb h') \sim  -\frac{1}{\sigma^2}\norm{\mb h' - \mb A_0\vec{\beta}}_2^2 - \vec{\beta}^\H\mb E^{-1}\vec{\beta}.
\end{equation}
We can find that the problem \eqref{eq: loss} is to minimize the posterior with respect to $(\vec{\tau}, \theta,\vec{\beta})$. So our QNOMP method is going to find the maximum a posterior (MAP) estimation of $(\vec{\tau},\vec{\beta})$. The regularizer $\mb E$ plays the role of prior covariance of $\vec{\beta}$. The extrapolations on $\{f_1,\dots,f_K\}$ are $\tilde{\mb h} = \mb A_e(\vec{\tau}^{\text{MAP}})\hat{\vec{\beta}}$.

MAP is a widely used estimator but may not be optimal. Measuring the error by mean squared error, it is well-known  \cite{Kay93CRB} that the minimum mean squared error estimator (MMSE) is the posterior mean, i.e., $\vec{\tau}^{\text{MMSE}}=\mathbb{E}_{\vec{\tau}|\mb h'}\vec{\tau}$, $\vec{\beta}^{\text{MMSE}}=\mathbb{E}_{\vec{\beta}|\mb h'}\vec{\beta}$. The corresponding extrapolated CSI is
    \begin{equation}\label{eq:mmse}
      \tilde{\mb h}= \mathbb{E}_{\vec{\tau}|\mb h'}\mb A_e(\bd \tau)\mathbb{E}(\bd \beta|\bd\tau,\mb h'),
    \end{equation}
    where the conditional posterior mean $\mathbb{E}(\bd \beta|\bd\tau,\mb h')=(\mb A_0^\H \mb A_0+\sigma^2\mb E^{-1})^{-1}\mb A_0^\H\mb h'$, which is exactly \eqref{eq:regbeta}. If the posterior distribution of $\vec{\tau}| \mb h'$ is $\delta(\vec{\tau}-\vec{\tau}^{\text{MAP}})$, the MMSE coincides with the MAP, which can be calculated by QNOMP.

    \subsection{Linear Optimal Extrapolation}\label{Sec:4A}

The exact calculation of MMSE can be very difficult, even if the posterior is given. Instead, we can use the linear minimum mean squared error estimator (LMMSE) for the extrapolated CSI $\tilde{\mb h}$. Namely, we are going to find the solution of the following problem
\begin{equation}\label{eq:plmmse}
  \min_{A,b}\ \mathbb{E}\big\|\tilde{\mb h} - \mb A_e(\vec{\tau})\vec{\beta}\big\|_2^2,\ \ \text{s.t. }  \tilde{\mb h}=A\mb h' +b.
\end{equation}
    We can show that the solution to this problem is
    \begin{equation}\label{eq:lmmse}
    \begin{aligned}
      \tilde{\mb h} & = \cov(\mb A_e\vec{\beta}, \mb h')\cov(\mb h', \mb h')^{-1}\mb h'\\
                    & = \cov(\mb A_e\vec{\beta}, \mb A_0\vec{\beta})(\cov(\mb A_0\vec{\beta}, \mb A_0\vec{\beta})+\sigma^2 I)^{-1}\mb h',
    \end{aligned}
    \end{equation}
     where $\cov(\vec{a}, \vec{b})=\mathbb{E}_{\vec{\tau}|\mb h'}\mathbb{E}_{\vec{\beta}\sim CN(0,\mb E)}\vec{a}\vec{b}^\H$ is the covariance matrix of random vectors $\vec{a}$ and $\vec{b}$.

The LMMSE is a good approximation to the optimal MMSE estimator. In fact, we have
\begin{equation}\label{eq:appMMSE}
\begin{aligned}
\tilde{\mb h}&=\mathbb{E}_{\bd\tau|\mb h'} \mb A_e(\mb A_0^\H\mb A_0+\sigma^2\mb E^{-1})^{-1}\mb A_0^\H \mb h'\\
&=\mathbb{E}_{\bd\tau|\mb h'} \mb A_e\mb E\mb A_0^\H(\mb A_0\mb E\mb A_0^\H+\sigma^2 I)^{-1}\mb h'.\\
\end{aligned}
\end{equation}
    If we make the approximation 
     \[\begin{aligned}\mathbb{E}_{\bd\tau|\mb h'} &\mb A_e\mb E\mb A_0^\H(\mb A_0\mb E\mb A_0^\H+\sigma^2 I)^{-1}\\
     &\approx\mathbb{E}_{\bd\tau|\mb h'} \mb A_e\mb E\mb A_0^\H(\mathbb{E}_{\bd\tau|\mb h'}\mb A_0\mb E\mb A_0^\H+\sigma^2 I)^{-1},\end{aligned}\]
     the MMSE can be approximated by
     \begin{equation}\label{eq:appMMSE2}
     \begin{aligned}
       \tilde{\mb h}&\approx \mathbb{E}_{\bd\tau|\mb h'} \mb A_e\mb E\mb A_0^\H(\mathbb{E}_{\bd\tau|\mb h'}\mb A_0\mb E\mb A_0^\H+\sigma^2 I)^{-1}\mb h'\\
&=\mathrm{Cov}(\mb A_e\bd \beta,\mb A_0\bd \beta)(\mathrm{Cov}(\mb A_0\bd \beta)+\sigma^2  I)^{-1}\mb h',
\end{aligned}
\end{equation}
which is exactly the LMMSE \eqref{eq:lmmse}. So LMMSE is an explicit approximation of MMSE. Let us call it the linear optimal extrapolation (LOX).

To compute the LOX \eqref{eq:lmmse}, one has to compute the expectation with respect to $\vec{\tau}|\mb h'$. Further assuming the entries of $\vec{\beta}$ are uncorrelated, $\mb E$ has a diagonal form $\mb E=\text{diag}(E_1, E_2,\dots, E_{N_p})$, we have
    \begin{equation}\label{eq:compcov}
    \begin{aligned}
 \mathrm{Cov}(\mb A_0\bd \beta, \mb A_0\bd \beta)&=\mathbb{E}_{\bd\tau|\mb h'}\mb A_0\mb E\mb A_0^\H\\
 &=\sum_{i=1}^{N_p} \mathbb{E}_{\tau_i|\mb h'}  E_i\mb a_0(\tau_i)\mb a_0^\H(\tau_i)
    \end{aligned}
    \end{equation}
    and
    \begin{equation}
    \begin{aligned}
 \mathrm{Cov}(\mb A_e\bd \beta, \mb A_0\bd \beta)&=\mathbb{E}_{\bd\tau|\mb h'}\mb A_e\mb E\mb A_0^\H\\
 &=\sum_{i=1}^{N_p} \mathbb{E}_{\tau_i|\mb h'}  E_i\mb a_e(\tau_i)\mb a_0^\H(\tau_i).
    \end{aligned}
    \end{equation}
So the expectation in high dimensions can be reduced to $N_p$ one dimensional expectations. To further compute $\mathbb{E}_{\vec{\tau}|\mb h'}(\cdot)$, we utilize the approximation that the distribution of $\vec{\tau}|\mb h'$ is roughly a Gaussian distribution $N(\bd \tau|\bd \tau^{\text{MAP}},H^{-1})$, where $H$ is the Hessian of function \eqref{eq:optim} (see Appendix \ref{sec:app2} for heuristic arguments). Therefore, $\tau_i|\mb h'$ is normally distributed with variance $(H)^{-1}_{ii}$. Thanks to QNOMP, in the BFGS step, the matrix $H_n$ is the approximation of $H^{-1}$. So we can  get a Gaussian approximation of the posterior automatically without additional computation:
 \[\bd \tau|\mb h'\approx N(\bd \tau|\bd \tau^{\text{MAP}},H_n).\]
We can approximate the expectation by Gaussian quadrature. To be specific, we have
\[\begin{aligned}
\mathbb{E}_{\tau_i|\mb h'}  & \mb a_0(\tau_i)\mb a_0(\tau_i)^\H \approx \sum_{k=1}^{S} \mb a_0(\tau_i^k)\mb a_0^\H(\tau_i^k) w_k,\\
\mathbb{E}_{\tau_i|\mb h'}  & \mb a_e(\tau_i)\mb a_0(\tau_i)^\H \approx \sum_{k=1}^{S} \mb a_e(\tau_i^k)\mb a_0^\H(\tau_i^k) w_k,
\end{aligned}\]
where $S$ is the order of Gaussian quadrature, $\tau_i^k = \tau_i^{\text{MAP}}+\sqrt{2\delta_i}x_k$, $\delta_i = 1/(H_n)_{ii}$, $x_k$ are Gaussian quadrature points and $w_k$ the corresponding quadrature weights, which can be pre-computed. Thus the covariance can be approximated by
      \begin{equation}\label{eq:gausscov}
        \begin{aligned}
        \mathrm{Cov}(\mb A_0\bd \beta, \mb A_0\bd \beta) & \approx \sum_{i=1}^{N_p}\sum_{k=1}^{S}E_iw_i\mb a_0(\tau_i^k)\mb a_0^\H(\tau_i^k)\\
        &= \mb B_0(\vec{\tau}_s)\mb D \mb B_0^\H(\vec{\tau}_s), \\
        \mathrm{Cov}(\mb A_e\bd \beta, \mb A_0\bd \beta) & \approx \sum_{i=1}^{N_p}\sum_{k=1}^{S}E_iw_i\mb a_e(\tau_i^k)\mb a_0^\H(\tau_i^k) \\
        &= \mb B_e(\vec{\tau}_s)\mb D \mb B_0^\H(\vec{\tau}_s),
        \end{aligned}
      \end{equation}
      where $\vec{\tau}_s=[\tau_1^1,\dots, \tau_1^S, \dots \tau_{N_p}^1,\dots,\tau_{N_p}^S]\in\mathbb{R}^{SN_p}$, $\mb B_0=\mb A_0(\vec{\tau}_s)\in \mathbb{C}^{M\times SN_p}$, $\mb B_e=\mb A_e(\vec{\tau}_s)\in\mathbb{C}^{K\times SN_p}$, $\mb D=\text{diag}\{{[E_1^b\dots E_{N_p}^b]\otimes \vec{w}}\}$, and $\vec{w}=[w_1,\dots , w_S]$. Thus the LOX has the form
\begin{equation}\label{eq:lmmsemat}
\begin{aligned}
\tilde{\mb h} &\approx \mb B_e \mb D \mb B_0^\H(\mb B_0 \mb D \mb B_0^\H+\sigma^2  I)^{-1}\mb h'\\
 &=\mb B_e(\mb B_0^\H\mb B_0+\sigma^2 \mb D^{-1})^{-1}\mb B_0^\H\mb h'.
\end{aligned}
\end{equation}
 Since Gaussian quadrature is very accurate, the order $S$ needs not to be large. In practice, we find that three points are already enough. The corresponding Gauss points are $x_{1,2,3}=-\sqrt{3/2}, 0, \sqrt{3/2}$, and weights $w_{1,2,3}=1/4,1/2,1/4$.

\subsection{Optimal Low Rank Approximation of LOX} \label{Sec:4B}

The direct computation of LOX involves the covariance computation and matrix inverse, which is expensive. We formulate a low rank approximation of LOX with the form
\begin{equation}\label{eq:flmmse}
\begin{aligned}
\min_{U,b}&\, \mathbb{E}\norm{\tilde{\mb h} - \mb A_e(\vec{\tau})\vec{\beta}}_2^2,\ \ \text{s.t.} & \tilde{\mb h}=U(\mb A_0\mb h' +b),
\end{aligned}
\end{equation}
where the columns of matrix $U=[u_1,u_2,\dots, u_r]\in\mathbb{C}^{M\times r}$ is a set of basis, $r$ is the rank. The solution of \eqref{eq:flmmse} is
\begin{equation}\label{eq:lowranklmmse}
Z = \mathcal{P}_U\cdot \cov(\mb A_e(\vec{\tau})\vec{\beta}, \mb h')\cov(\mb h', \mb h')^{-1}\mb h'.
\end{equation}
where $\mathcal{P}_U:=U(U^\H U)^{-1}U^\H$. This is actually the orthogonal projection of LOX \eqref{eq:lmmse} onto the column space of $U$. Without loss of generality, we can assume the basis are orthogonal, which means $U^\H U=I$. The MSE of estimator $Z$ is $\mathbb{E} Z^\H Z$, which has the form
    \begin{equation}\label{eq:mse}
\mathrm{Tr}[U^\H\cov(Z,\mb h')\cov(\mb h', \mb h')^{-1}\cov(\mb h',Z)U].
    \end{equation}
Thus, we can choose orthogonal basis $U$ properly to minimize the MSE. It is well known that the basis that minimizing MSE is the first $r$ eigenvectors of $\cov(Z,\mb h')\cov(\mb h', \mb h')^{-1}\cov(\mb h',Z)$. Moreover, if $Z-\mb h'\sim CN(0,\sigma^2I)$, $\{u_1,\dots, u_r\}$ can be chosen to be the first $r$ eigenvectors of $\cov(\mb h', \mb h')=\cov(\mb A_0\vec{\beta}, \mb A_0\vec{\beta})+\sigma^2I$.

    We point out that the optimal basis for low rank LOX is closely related to the discrete prolate spheroidal sequences (DPSS) or discrete Slepian basis \cite{DPSS}. In fact, we show that when $N_p=1$, the eigenvectors $\{u_1,u_2,\dots, u_r\}$ are exactly DPSS. Details are discussed in Appendix \ref{sec:app1}.

\subsection{The Block Reweighting Technique}\label{Sec:4C}

In realistic communications, the propagation paths are not distributed completely in random, but rather has a clustered structure \cite{Tse2005}. The multipath model $\mb h_k$ in \eqref{eq: multipath} can be rearranged as
\[\mb h_k=\sum_{i=1}^{N_c} e^{-\j 2\pi \tau_i k\Delta f}\sum_{j=1}^{N_i} \beta_{i,j}   \mb{a}_\theta\left(\theta_{i,j}\right),\]
where $N_c$ is the number of scattering clusters, $N_i$ is the number of sub-paths in the $i$-th scattering cluster. Multiple sub-paths in a path cluster have the same or very close time delays, yet the spatial angles of the sub-paths concentrate in a certain extended range, which is referred to as the angular spread of the path cluster. We denote the angular blocks as $C_i=\{\theta_{i,j}|j=1,2,\dots, N_i\}, i=1,2,\dots, N_c$.  Each angular block is assumed to be concentrated with an angular spread:
\begin{equation}\label{eq: block}
C_i\subset B_i=(\theta_{i,0}-\Delta_i,\theta_{i,0}+\Delta_i).
\end{equation}

Taking advantage of this block sparsity structure of channel, we can improve QNOMP to get a more accurate result. After the QNOMP, we have got the estimated parameters $(\bd \tau,\bd \theta,\bd \beta)$ for major propagation paths. These paths can be viewed as the centers of the angular blocks $B_i$ in \eqref{eq: block}. The path parameter $\tau_i$ is the delay of the cluster, $\theta_i$  the center of the angular block, and $|\beta_i|^2$  approximately the total energy of the angular block. Here we propose a simple algorithm to adaptively reconstruct the angular blocks and calculate the final channel parameters based on this prior structure.

    Given $(\bd \tau,\bd \theta,\bd \beta)$, for every $(\tau_i, \theta_i, \beta_i)$, we expand it as an angular block $\{(\tau_i,\theta_i)|i=1,\cdots,N_p\}$ with $\Delta=\gamma\Delta_{\theta}$ as the maximum width. $\Delta_\theta$ is a unit width, $\gamma\geq 1$. To be specific, the block has $2\gamma+1$ sub-paths
    \begin{equation}\label{eq:block}
      B_i=\{(\tau_i,\theta_i+j\Delta_{\theta})|j=-\gamma:\gamma\}.
    \end{equation}
The $2\gamma+1$ paths have a total energy $E_i=\abs{\beta_i}^2$. We assume the energy is uniformly distributed on sub-paths with non-informative prior. Namely, the path $(\tau_i,\theta_i + j\Delta_\theta)$ carries energy
    \begin{equation}\label{eq:eqenergy}
      E_{ij}=\frac{|\beta_i|^2}{2\gamma+1}.
    \end{equation}

    We denote all the sub-paths in blocks as $(\bd \tau^b,\bd \theta^b,\mb E^b)=\mathcal{V}\{(\tau_{ij},\theta_{ij},E_{ij})=(\tau_i,\theta_i+j\Delta_{\theta},E_{ij})\big|i=1:N_p,j=-\gamma:\gamma\}$.
     Then a regularized LMMSE is utilized to estimate the energy of each sub-path:
\begin{equation}\label{eq: blmmse1}
\bd h(\mb E^b)=(\mb A_b^\H\mb A_b+\sigma^2  (\mb E^b)^{-1})^{-1}\mb A_b^\H \mb h',
\end{equation}
where $\mb A_b=\mb A(\bd \tau^b,\bd \theta^b)$. Thus $\bd h(\mb E^b)$ is the channel gains for each sub-path under the uniform energy $\mb E^b$ and Gaussian prior. The energy of the $i$-th sub-path is $\abs{h_i(\mb E^b)}^2$. So we can change the prior $\mb E^b$ to $\mb E_{\bd h}=\text{diag}\{\abs{h_1(\mb E)}^2, \abs{h_2(\mb E)}^2,\dots\}$ and re-estimate the channel gains by
\begin{equation}\label{eq:final}
\bd h^b=(\mb A_b^\H\mb A_b+\sigma^2 \mb E_{\bd h}^{-1})^{-1}\mb A_b^\H \mb h'.
\end{equation}
Finally $(\bd \tau^b,\bd \theta^b,\bd h^b)$ gives the parameters for all sub-paths. Although this reweighting technique can be applied recurrently, we find practically it is enough to reweight only once.

The total number of sub-paths can reach $(2K+1)N_p$. The main cost of block reweighting lies in  \eqref{eq: blmmse1} and \eqref{eq:final}, which is $O(((2k+1)N_p)^3+((2k+1)N_p)^2(M+N)+N_pMN)$. In practice, the block reweighting introduces a slightly larger computational cost. However, by making use of the prior knowledge of the channel structure, this procedure can further identity low energy sub-paths that are closely distributed near the major paths, which increases the overall accuracy. We summarize the block reweighting technique in Algorithm 4.
\begin{algorithm}\label{alg:4}
\caption{Block Reweighting Technique}
\begin{algorithmic}[1]
\State \textbf{Input:} Block center and energy $(\vec{\tau},\vec{\theta},\vec{\beta})$ obtained by QNOMP, CSI $\mb h'$, angle spread $\gamma\Delta_\theta$, threshold $\epsilon, \epsilon_1, \epsilon_2$, noise level $\sigma$,
\State \textbf{Output:} Sub-path parameters $(\bd \tau^b,\bd \theta^b,\bd h^b)$
        \State Divide $|\beta_i|^2,i\in [N_p]$ into two parts $[N_p]=I_H\cup I_L$ by selecting the first few paths that $\sum_{i\in I_H} |\beta_i|^2>(1-\epsilon) \norm{\vec{\beta}}^2$
\State Construct angular blocks $B_i$ for $i\in I_H$ by \eqref{eq:block} and initialize a uniform energy distribution \eqref{eq:eqenergy}
        \State The total sub-paths are $(\vec{\tau}^b,\vec{\theta}^b,\mb E^b)=\cup_i (B_i, E_i)\cup_{i\in I_L} (\tau_i,\theta_i, \abs{\beta_i}^2)$
        \State Apply LMMSE \eqref{eq: blmmse1} to re-estimate $\mb E_{\mb h}$
        \State Apply LMMSE \eqref{eq:final} to have the estimation $\mb h^b$\\
        \Return $(\bd \tau^b,\bd \theta^b,\bd h^b)$
\end{algorithmic}
\end{algorithm}

\section{Numerical Results}

We conduct various numerical tests of QNOMP on both simulated models and CDL-C models. We compare the extrapolation accuracy and the computational cost for different related methods in numerical experiments. All the tested cases are TDD  MIMO channels. The pilots are carried on $M=24$ subcarriers with spacing $\Delta f=240$KHz. The BS is equipped with a ULA with $N=64$ antenna elements, while the user has one antenna.

We take the conventional OMP method, NOMP and a dynamic off-grid ISTA method (OG-ISTA) \cite{NonUniform} with gradient descent based updating to make the comparison. In OG-ISTA, a compressing grid technique \cite{chen2021application} is also applied to reduce the computational cost. QNOMP and LOX are compared with them. The QNOMP with block reweighting technique is also tested in the clustered channel scenario. To make a fair comparison, the grid of OMP and NOMP is set to be the same as the finest resolution of local  refinement (i.e., the multi-resolution OMP) module in QNOMP.

We set a $K$-times frequency bandwidth to test the performance of extrapolation. More specifically, the pilot band frequencies are $f_0+\Delta f\cdot [0,1,\dots,M-1]$. The band frequencies to extrapolate are $f_0+\Delta f\cdot[kM,kM+1,\cdots,(k+1)M-1]$ for $k=1:K$. This is actually a one-sided extrapolation.

For the same fixed ground truth $\mb h$, let AWGN $\mb w\sim CN(0,\sigma^2  I_{NM}),\mb h'=\mb h+\mb w$ as the noisy channel. The SNR is defined as $\text{SNR}:=\|\mb h\|_2^2/(\sigma^2 NM)$. Under different noise level $\sigma$, 100 simulations are carried out, and the parameters $(\bd \tau,\bd \theta)$ are sampled from corresponding uniform distributions in each simulation. The accuracy of the estimated $\mb{\hat{h}}$ is measured by the average error for the 100 simulations, i.e.,
\[\text{NMSE}=\frac{1}{100}\sum_{i=1}^{100}\frac{\|\mb h_i-\mb{\hat{h}}_i\|_2^2}{\|\mb h_i\|_2^2}. \]
The SNR is also averaged over simulations.

For simulated channel models, we also measure the super-resolution error for the channel parameters in virtual domain. Without loss of generality, we only consider the delay $\vec{\tau}$. The average estimation error is measured by
\[{\text{NMSE}}=\frac{1}{100}\sum_{i=1}^{100}\frac{\|\bd \tau_i-\bd {\hat\tau}_i\|_2^2}{\mathcal{L}(\Delta_{\text{dft}}^{\tau})^2}, \]
where the $k$-th component of the estimated delay $\bd {\hat\tau}_i$ is the one that closest to the true delay $\tau_{i,k}$. The estimation error is also compared with the Cramer-Rao lower bound \cite{Kay93CRB} (CRB). All experiments are carried out in Matlab R2022b, on a i5-13500H PC. The algorithm's time consumption is measured in CPU time.

\subsection{Simulated Multipath Propagation Scenario}\label{sec:multipath}

We construct a simple sparse multipath propagation model to test the performance of different algorithms. In this model, there are 7 paths with corresponding delays and angles $(\tau_k,\theta_k)$ for $k=1:7$. These delays are set to be equally spaced $\tau_{k+1}-\tau_k=C_1\Delta_{\text{dft}}^\tau,\theta_{k+1}-\theta_k=C_2\Delta_{\text{dft}}^\theta$. The constants $C_{1,2}$ indicates the proximity of paths. The smaller the constants are, the closer the paths lie. The channel gains $\bd \beta=e^{\j 2\pi \bd \phi}$ with $\phi_k \stackrel{\text{i.i.d.}}{\sim} \text{Uniform}[0,1]$.  The average NMSEs for 100 simulations are calculated  under SNR $\approx 8.5$ dB.

The local refinement rates of QNOMP is set to be $k_1=k_2=10$ and refinement times $n^{\text{LR}}=1$. The number of BFGS iterations are set as $n^{\text{in}}=3$ and $n^{\text{out}}$ undetermined until convergence. The iteration parameters in NOMP are set as $R_s=1,R_c=3$ and we add an $n^{\text{out}}$ times cyclic optimization as in QNOMP. The grid size of conventional OMP and NOMP is $0.1\Delta_{\text{dft}}^{\tau,\theta}$.

\begin{figure*}[!htbp]
\centering
\subfloat[Estimation and extrapolation errors for Scenario 1.]{\includegraphics[width=.9\columnwidth]{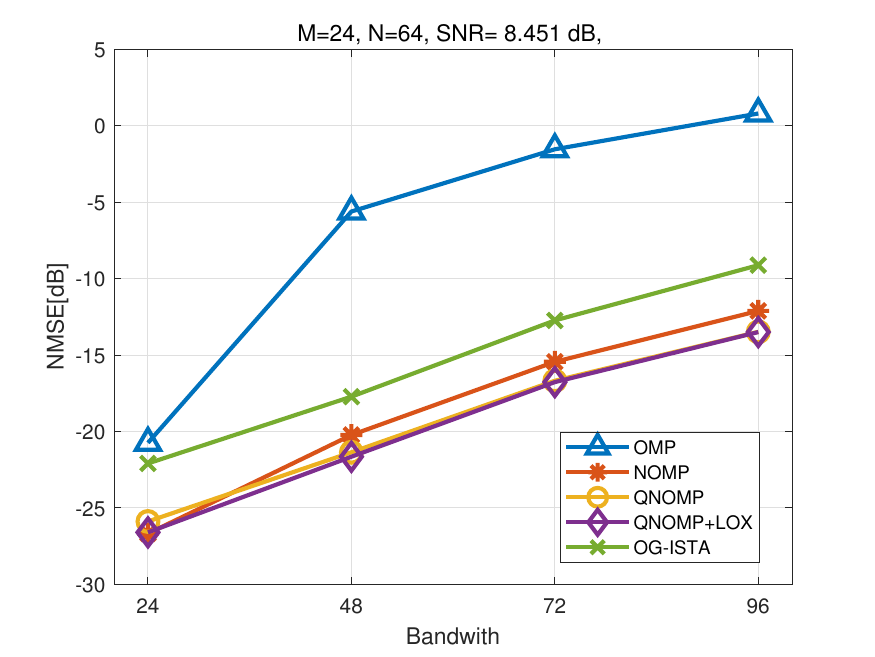}
\label{pica}
}
\subfloat[Estimation errors of delay for Scenario 1.]{\includegraphics[width=.9\columnwidth]{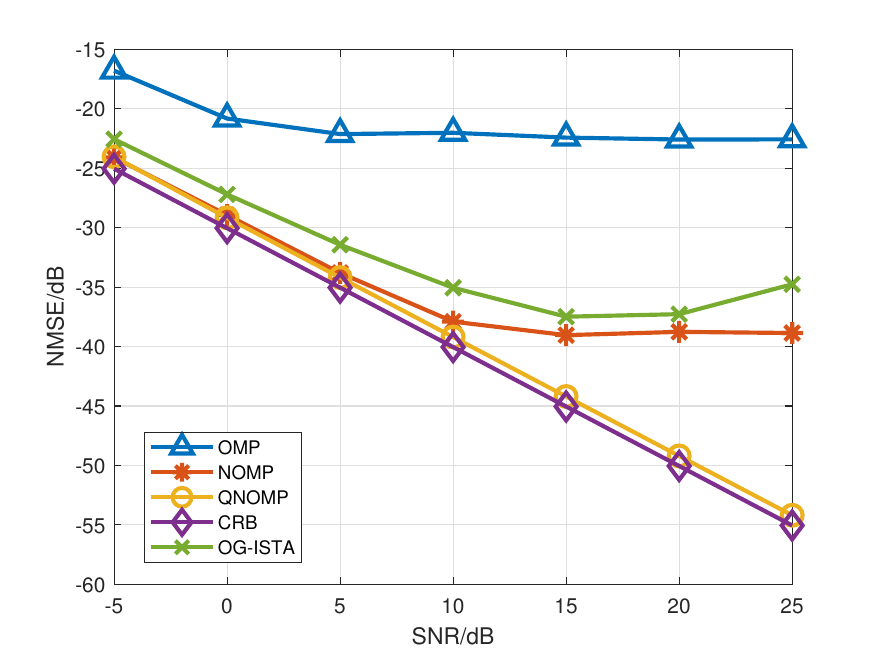}
\label{picb}
}
\\
\subfloat[Estimation and extrapolation errors for Scenario 2.]{\includegraphics[width=.9\columnwidth]{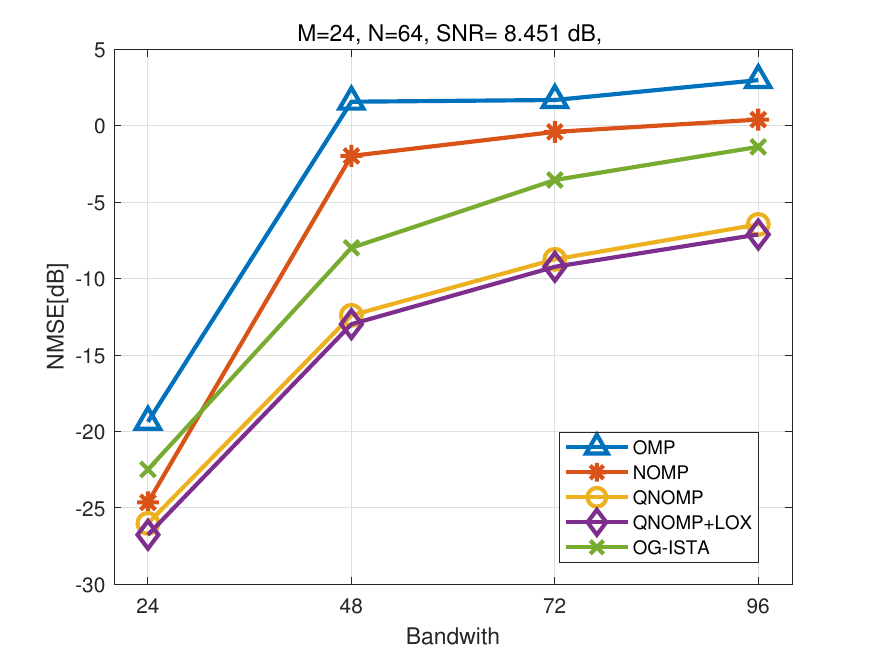}
\label{picc}
}
\subfloat[Estimation errors of delay for Scenario 2.]{\includegraphics[width=.9\columnwidth]{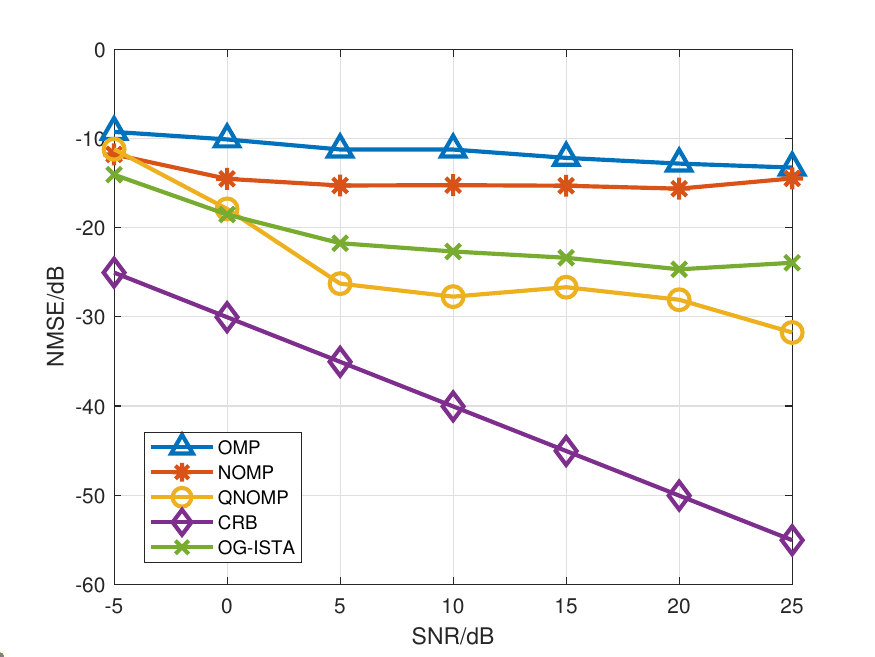}
\label{picd}
}

\caption{Estimation and extrapolation error of different algorithms for two typical scenarios and the super-resolution error for virtual delay variable in simulated channel. Scenario 1 has a sparse delay distribution with $C_1=2,C_2=0.5$. Scenario 2 has a dense delay distribution with $C_1=1,C_2=0.5$. (a),(c): Estimation and extrapolation error for Scenarios 1, 2. The bandwidth with label 24 corresponds to the estimation error. The channel is extrapolated to the bandwidth with label 48, 72 and 96 ($K=2,3,4$). (b)(d): Estimation error of delay for Scenarios 1 and 2. }
\label{fig:multipath}
\end{figure*}

The numerical experiments are carried out under two different  scenarios which have different path distributions. In Scenario 1, we set $C_1=2$, $C_2=0.5$, which has a relatively sparse path-delay distribution. In Scenario 2, parameters are set as $C_1=1, C_2=0.5$, which has a dense path-delay distribution. We present the NMSE of CSI and delay for different algorithms under different scenarios in Fig.~\ref{fig:multipath}. The bandwidth with the label 24 is actually the result of CSI estimation, and larger label numbers correspond to the CSI extrapolation. It can be observed in Fig.~\ref{fig:multipath}(a)(c) that the QNOMP and the LOX based on QNOMP perform similarly and better than the others. As  the extrapolated bandwidth is wider, the extrapolation error of all algorithms increases. The two Newton type methods NOMP and QNOMP outperform the others in general. It is worth noting that although NOMP has a similar estimation error as QNOMP, the QNOMP has better extrapolation performance. Furthermore, the extrapolation error of QNOMP and LOX are not only lower, but also increase more slowly. This is due to the better estimation of super-resolution delay parameters, as shown in Fig.~\ref{fig:multipath}(b)(d). So, no matter the paths are concentratively or scatteredly distributed, QNOMP shows significant gains on both CSI estimation and extrapolation compared to conventional OMP and NOMP. When the paths are not dense, the error of QNOMP can achieve CRB for wide range of SNR. The LOX also provides an improvement over the two-stage QNOMP as well.

\subsection{Simulated Channel with angular block Structure}

To show the improvement of block reweighting technique, we construct a clustered channel model. There are three clusters with corresponding delays $\tau_1,\tau_2,\tau_3$ and angular centers $\theta_1, \theta_2, \theta_3$. In each cluster, there are corresponding common angular blocks $B_1,B_2,B_3$, where $B_i=\{\theta_i+jC_2\Delta_{\text{dft}}^{\theta}|j=-2:2\}$. The gains of sub-paths are $\bd \beta=e^{\j 2\pi \bd \phi}$ with $\phi_i\stackrel{\text{i.i.d.}}{\sim} \text{Uniform}[0,1]$. We will test the CSI estimation performance ($K=1$) and the extrapolation performance for $K=2$.

    The settings of QNOMP and baseline OMP algorithm are the same as in Section \ref{sec:multipath}. The unit block width of block reweighting is $\Delta_{\theta}=0.5\Delta_{\text{dft}}^\theta$. The angle spread is set as $\gamma=4$. We skip the procedure of selecting high energy paths, i.e., $\epsilon=0$.
    We also test algorithms by designing different scenarios. In Scenario 1, we set $C_1=C_2=2$ indicating the case that paths distributes sparsely in both delay and angular domain. In Scenario 2, $C_1=1, C_2=2$, which has dense paths in delay domain. In Scenario 3, $C_1=1, C_2=0.5$, in which paths are dense in both delay and angular domain.

\begin{figure*}[htbp]
\centering
\subfloat[Scenario 1. $k=1$ Channel NMSE ]{\includegraphics[width=.7\columnwidth]{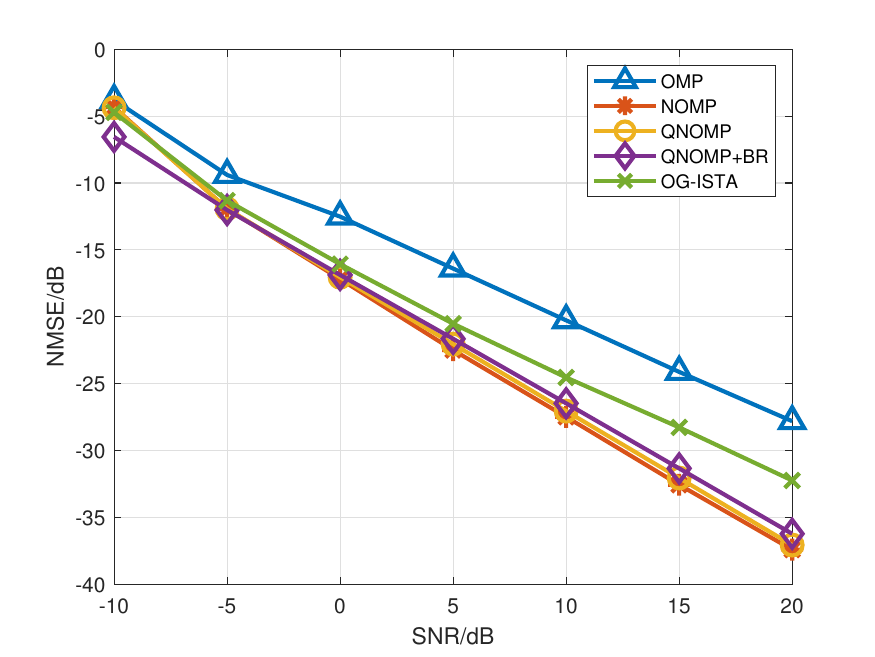}
}
\subfloat[Scenario 2. $k=1$ Channel NMSE ]{\includegraphics[width=.7\columnwidth]{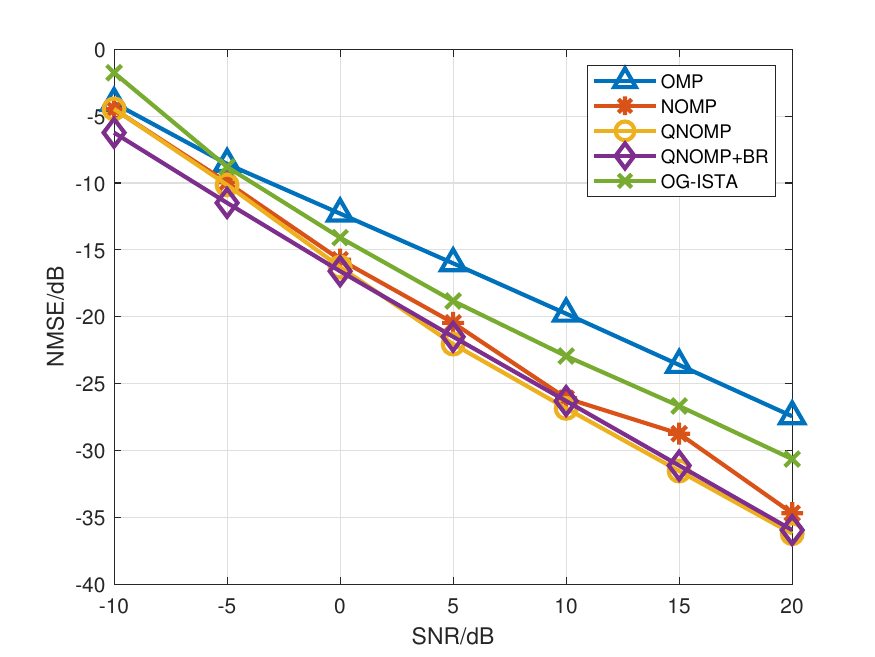}
}
\subfloat[Scenario 3. $k=1$ Channel NMSE]{\includegraphics[width=.7\columnwidth]{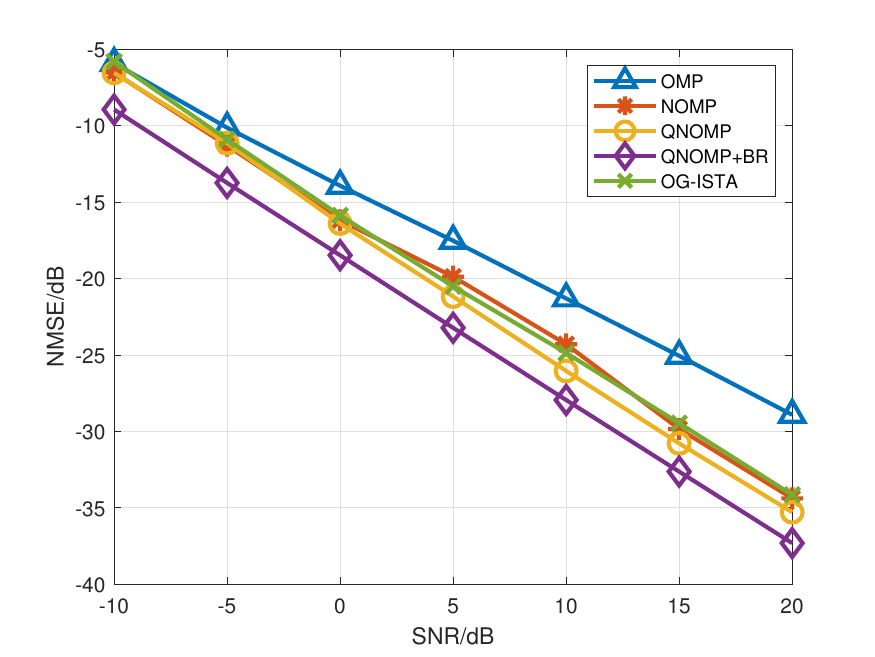}
}
        \\
\subfloat[Scenario 1. $k=2$ Channel NMSE ]{\includegraphics[width=.7\columnwidth]{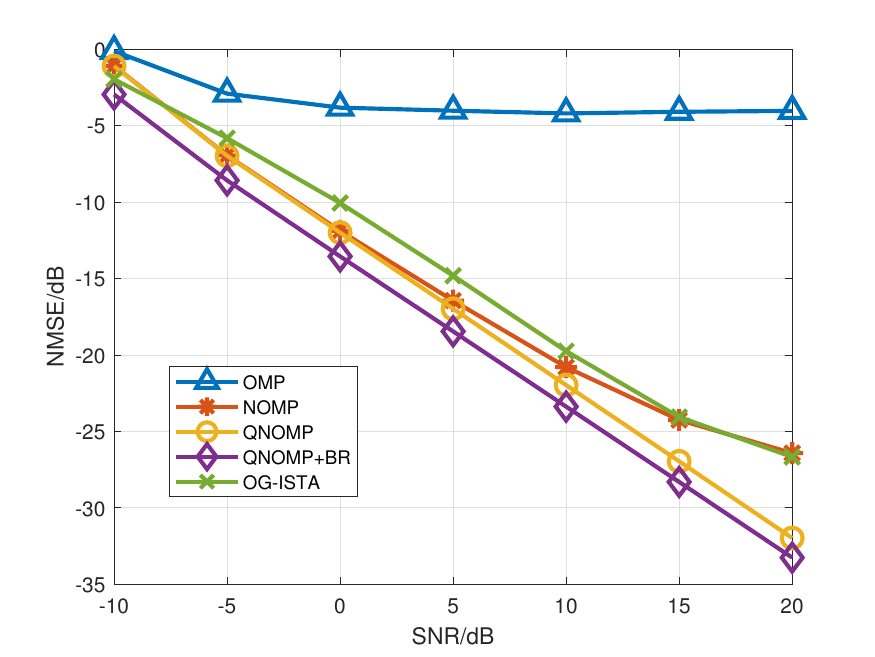}
}
\subfloat[Scenario 2. $k=2$ Channel NMSE]{\includegraphics[width=.7\columnwidth]{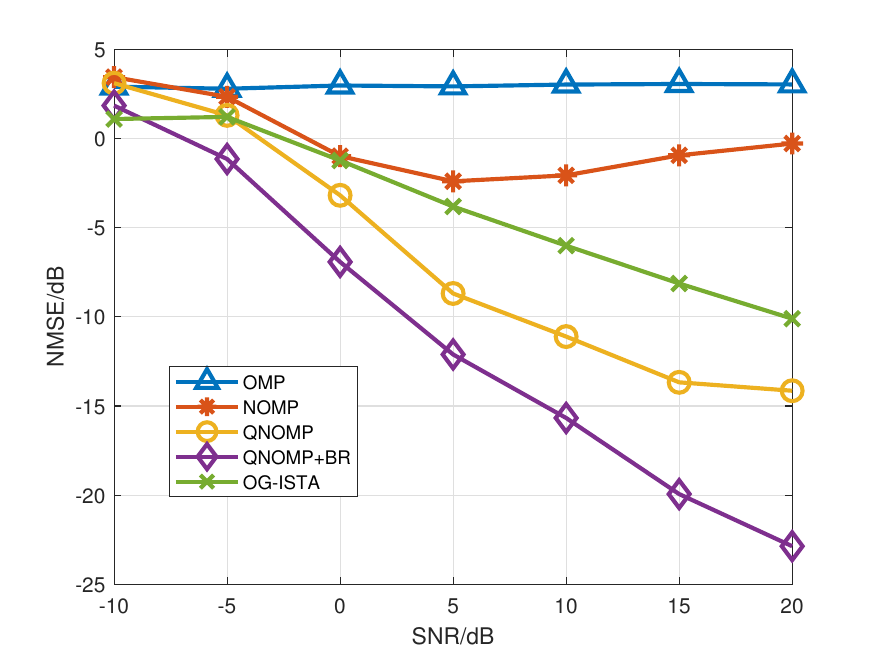}
}
\subfloat[Scenario 3. $k=2$ Channel NMSE]{\includegraphics[width=.7\columnwidth]{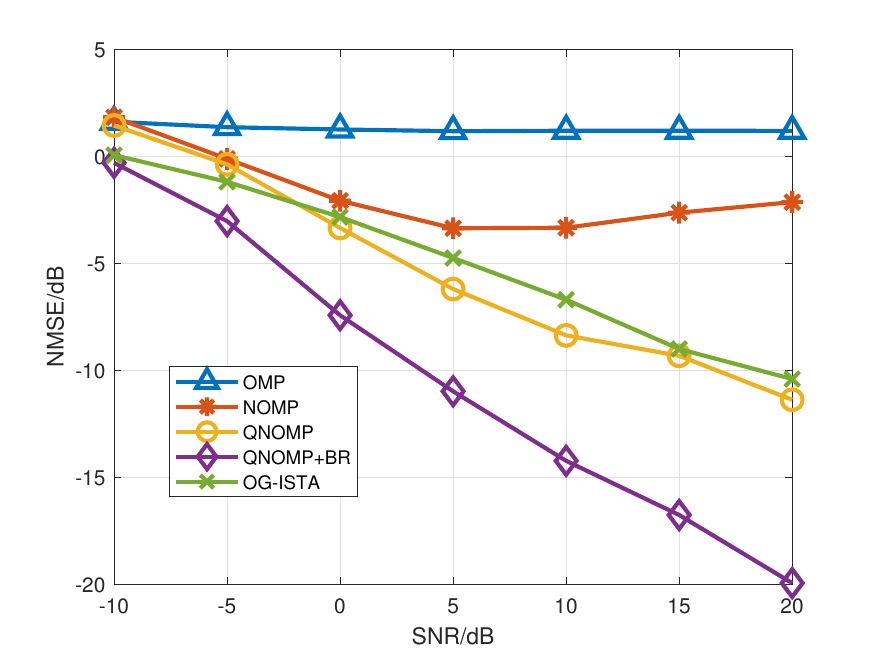}
}
\caption{Estimation and extrapolation error of different algorithms for three typical scenarios in clustered simulated channel. Scenario 1 has a sparse delay and angle distribution with $C_1=2, C_2=2$. Scenario 2 has a dense delay distribution and sparse angle distribution with $C_1=1,C_2=2$. Scenario 3 has a dense delay and angle distribution with $C_1=1, C_2=0.5$. (a)(b)(c): Estimation error with respect to (w.r.t.) SNR for Scenarios 1, 2, 3. (d)(e)(f): Extrapolation error w.r.t. SNR for Scenarios 1, 2, 3. The channel is extrapolated to bandwidth with label 48 ($K=2$).}
\label{fig1}
\end{figure*}

Fig. \ref{fig1} shows the channel estimation and extrapolation error of algorithms under different scenarios. It can be observed that QNOMP performs better in both estimation and extrapolation. Besides, in Scenario 3 which has both dense delay and angle, the block reweighting technique obtains a significant improvement over QNOMP and other algorithms by taking benefit of the angular block structure.

It is worth noting that the sparsity in delay domain does affect on the extrapolation performance. From Fig.~\ref{fig1}(d) and (e), one can find  that as $C_1$ reduces from 2 to 1, the extrapolation gets worse for both QNOMP with or without block reweighting. On the other hand, given $C_1$, the sparsity of angular block has little effect on extrapolation. This can be verified by comparing Fig.~\ref{fig1}(e) and (f). Since extrapolation is required in the frequency domain rather than space/antenna domain, the estimation accuracy of the time delay is more important than the angle. Therefore, the density of time delays poses a greater challenge for super-resolution and is the main factor affecting the accuracy of extrapolation. The pilot spreads in the entire antenna domain and does not require extrapolation, so the resolution in the angular domain has less impact on the accuracy of channel extrapolation.

    \begin{figure*}[htbp]
    \centering
    \subfloat[Scenario 1. Delay NMSE ]{\includegraphics[width=.7\columnwidth]{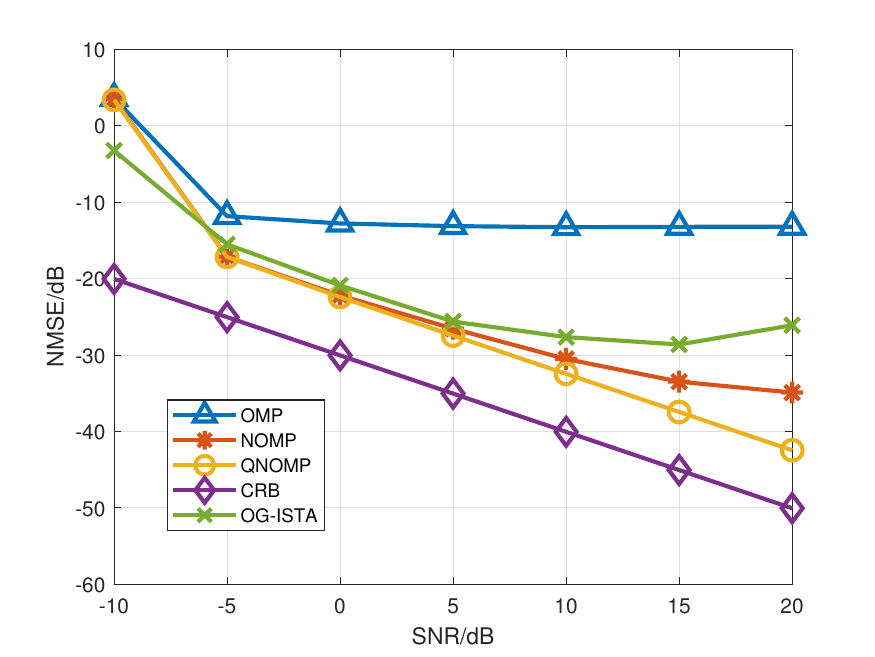}
}
\subfloat[Scenario 2. Delay NMSE]{\includegraphics[width=.7\columnwidth]{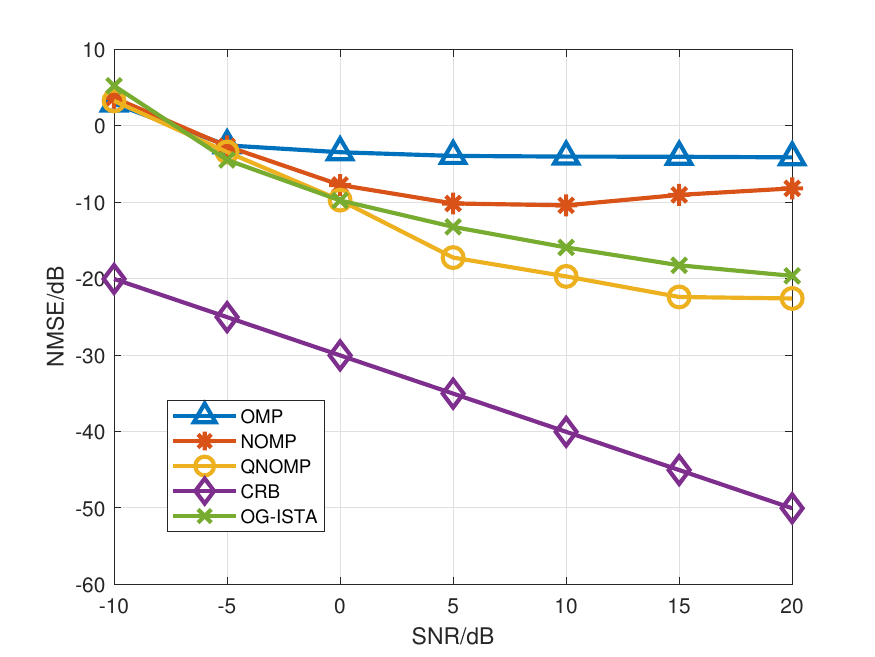}
}
\subfloat[Scenario 3. Delay NMSE]{\includegraphics[width=.7\columnwidth]{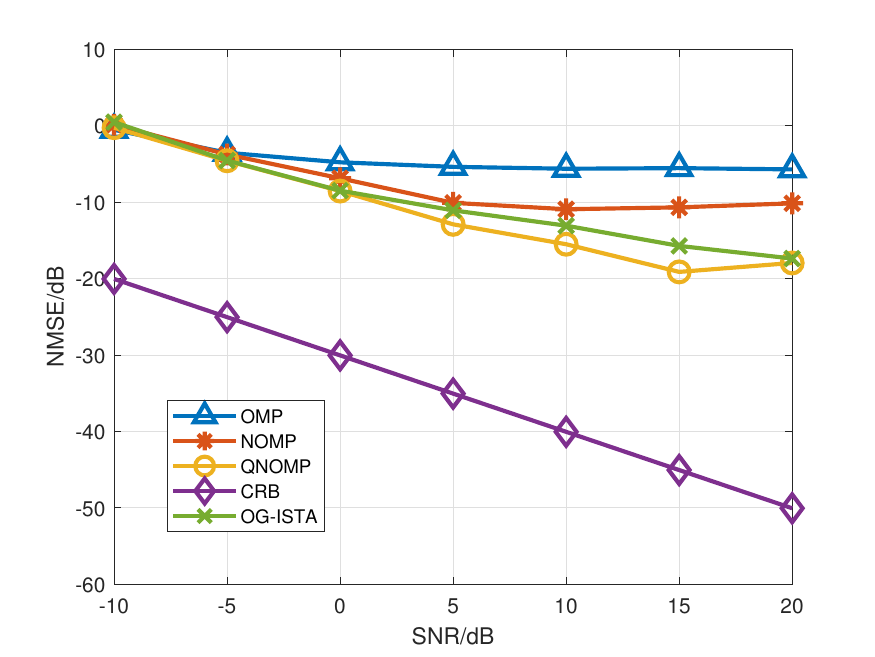}
}
\caption{Estimation error of virtual delay variable in clustered simulated channel. CRB is also plotted as a reference. The scenario setting is the same as in Fig. \ref{fig1}. (a)(b)(c): Estimation error of delay and CRB w.r.t. SNR for Scenarios 1, 2, 3.}
    \label{fig:crb}
    \end{figure*}

 We also examine the estimation error of the virtual delay $\vec{\tau}$ to verify the relation between delay estimation and extrapolation error. The estimation error of delay under different SNR is shown in Fig. \ref{fig:crb}, in which the CRB is also plotted as reference. It can be seen in Fig. \ref{fig:crb}(a) that as SNR increases, the delay estimation error of OMP stays at a almost constant level. Correspondingly, the extrapolation of OMP also stays at a constant level, as shown in Fig. \ref{fig1}. Similarly, the comparisons of Fig. \ref{fig:crb}(b)(c) and Fig. \ref{fig1}(e)(f) show that, as delay gets denser, OMP and NOMP cannot take benefit from the increase of SNR on estimating delay, so the extrapolations are also poor. However, the QNOMP still benefits from the increase of SNR so that the extrapolation error decreases. This suggests that QNOMP is a powerful algorithm for super-resolution and extrapolation.

\subsection{On CDL-C Model}

Simulations of the Clustered Delay Line (CDL) models in 3GPP \cite{3GPP} are performed using Matlab 5G NR Toolbox. The channel in this model has dense delay  clusters, and the angles on the clusters are concentrated within some angular blocks. The CDL-C model we used has a root mean square (RMS) time delay spreads of 100ns. The channel subcarrier spacing is 30kHz and the sampling interval is 60kHz.

The setting of QNOMP and NOMP iterations includes: the local refinement rate $k_1=k_2=10$, refinement times $n^{\text{LR}}=1$, the iteration numbers $n^{\text{in}}=3$ and $n^{\text{out}}=40$ for BFGS; $R_s=1,R_c=3, n^{\text{out}}=40$ for NOMP; and $n^{\text{it}}=200$ for the OG-ISTA iteration number. As a comparison, the grid size of conventional OMP is $0.1\Delta_{\text{dft}}^{\tau,\theta}$, which is the same as the finest resolution of QNOMP's on-grid module. The thresholds of block reweighting are $\epsilon=0.001$. The block width is $\Delta=6\Delta_\theta$, $\Delta_{\theta}=\Delta_{\text{dft}}^{\theta}$.

\begin{figure*}[!htbp]
\centering
\subfloat[Estimation error for CDL-C channel with 100ns RMS delay spread. ]{\includegraphics[width=.9\columnwidth]{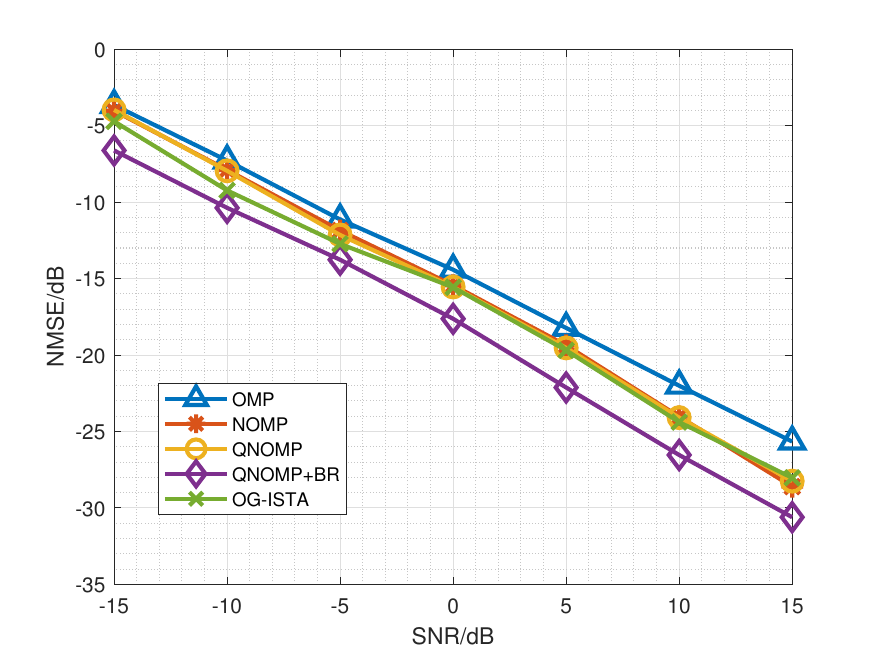}
\label{cdlc100-1}
}
\subfloat[Extrapolation error for CDL-C channel with 100ns RMS delay spread. The extrapolated bandwidth is $K=2$.]{\includegraphics[width=.9\columnwidth]{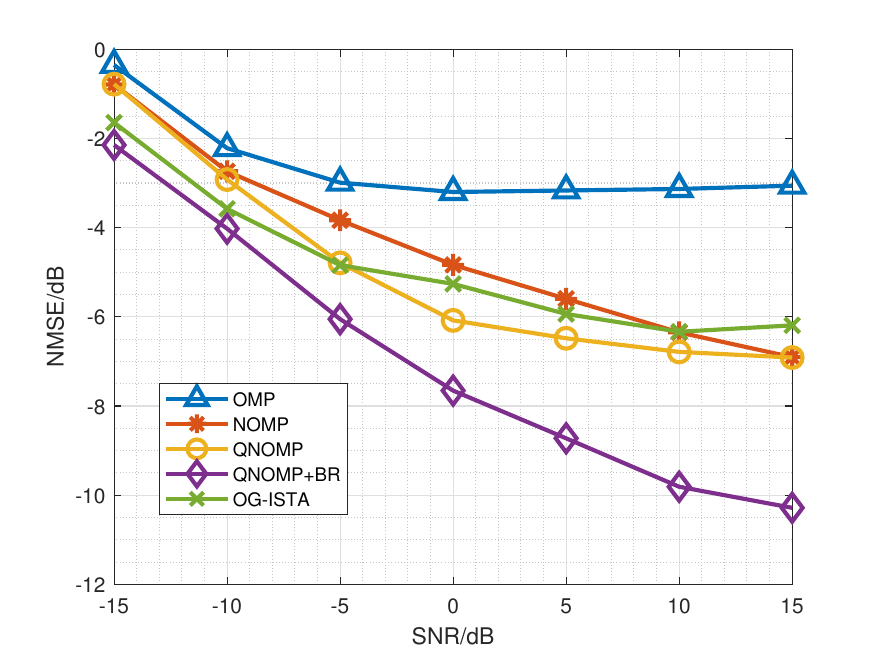}
\label{cdlc100-2}
}

\caption{Estimation and extrapolation error of different algorithms for CDL-C channel with RMS delay spread 100ns. }
\label{fig3}
\end{figure*}

    The estimation and extrapolation performance of different algorithms for CDL-C channel with 100ns RMS delay spread are shown in Fig. \ref{fig3}. Here we  take $M=96$ and $\Delta f=60$KHz. As in simulated channel cases, QNOMP significantly enhances the estimation and extrapolation performance. CDL-C channel is a typical model with angular blocks, so the block reweighting technique can greatly improve the performance of QNOMP for both low and high SNR by taking advantage of the prior clustered structure of the channel model.

\subsection{Computational Cost}\label{com}

We summarize the CPU time of different algorithms for CDL-C channel in Table~\ref{tab:2}. It can be found that the local refinement is more efficient than using a fine uniform grid. Comparing the OMP with grid size  $0.25\Delta_{\text{dft}}^{\tau,\theta}$ and OMP with local refinement ($k_1=k_2=10$), we find that the OMP with local refinement is 6-8 times faster and has a higher resolution. It is consistent with our theoretical analysis in Section \ref{sec:main}. For 2D or higher dimensional channels, the cost of computing residual dominates the overall cost of OMP, and the cost of computing residual is proportional to the number of paths $N_p$, vector sizes $M,N$ and refinement rates $k_1,k_2$. Thus using local refinement is highly beneficial.

    QNOMP is quite efficient in our test for CDL-C channel. To make a fair comparison, we also add the local refinement module in OMP and NOMP. It takes about 3-4 times CPU time of OMP in our setting, but achieves a significant estimation and extrapolation gain, as shown in Fig. \ref{fig3}. QNOMP takes even less time than NOMP and is more accurate. Moreover, it can also be observed that the cost of block reweighting is small. It only costs about 10\% more computing time than the QNOMP without block reweighting, but achieves a 2-5 dB gains in extrapolation.

We remark that since we keep the iteration number $n^{\text{it}}=200$ in OG-ISTA, the cost of it stays constant as the SNR increases. So OG-ISTA works better when finding an imprecise solution in very short time. It can catch a large part of the paths in a few iterations without matrix inversion.

\renewcommand{\arraystretch}{1.1}
\begin{table*}[htbp]
\begin{center}
\caption{CPU time (seconds) of algorithms for 100 simulations in CDL-C channel}\label{tab:2}
\begin{tabular}{c|c|c|c|c|c|c|c}
\hline\hline
\diagbox{Algorithm}{SNR/dB}
 &-15&-10&-5&0&5&10&15\\
\hline\hline
Conventional OMP ($0.25\Delta_{\text{dft}}$)&3.5394  &  4.0646  &  5.3405  &  7.6186  & 11.7528 &  18.6589 &24.7385\\
\hline
OMP (with local refinement) &0.4189   & 0.6551  &  0.6954   & 1.0064   & 1.4597    &2.0446    &2.9382\\
\hline
NOMP (with local refinement) & 1.7540 &   2.8609  &  4.9548  &  8.4036 &  14.1061  & 22.5925  & 32.9221\\
\hline
OG-ISTA (Compressing Grid)&9.9930  & 10.3610  & 10.6952 &  10.0479 &  10.0957   &10.0231 &  11.4636\\
\hline
QNOMP & 1.8053  &  1.9946  &  2.6501   & 3.6585  &  5.6925  &  8.2967  & 12.2837\\
\hline
QNOMP with Block Reweighting & {\bf 0.1741}  & {\bf  0.2164}  &  {\bf 0.2581} &  {\bf  0.3158} &  {\bf  0.6040}  &  {\bf 0.8387} &  {\bf  0.9337}\\
\hline\hline
\end{tabular}
\end{center}
\end{table*}

\section{Conclusion and Discussion}

We propose an efficient super-resolution algorithm named QNOMP for the CSI estimation and extrapolation in TDD MIMO system. The algorithm consists an on-grid sparse selection stage and an off-grid joint parameter optimization stage. In the on-grid stage, a local refinement module is introduced so the on-grid stage can achieve a high resolution with low cost. In the off-grid stage, we use the BFGS quasi-Newton optimization. The high efficiency of BFGS algorithm allows us to perform a joint optimization for all channel parameters. This joint optimization is shown to be crucial for improving the estimation and extrapolation accuracy. We also derive the linear optimal extrapolation, which further improves the extrapolation error. Furthermore, when the considered channel has angular block structure, the block reweighting technique takes benefit from this prior knowledge and leads to a better performance for both estimation and extrapolation.

QNOMP is simple and efficient. In the on-grid stage, we choose OMP for sparsity selection because  it has low computing complexity with local refinement. The joint optimization in off-grid stage is the critical step for a successful super-resolution. BFGS is one of the most efficient inverse free optimization method to achieve this goal. When $N_p$ is large, the limited-memory BFGS (L-BFGS) might be another candidate.

We also design a heuristic block reweighting module to take benefit of the angular block structure. Although far from being optimal, it does capture the intrinsic nature of considered channel and significantly improve the performance of super-resolution. Smart utilization of such prior information will be a universal topic in CSI estimation.

\appendices

\section{Gaussian Approximation of Posterior}\label{sec:app2}

We will first show that, in the high SNR regime, the posterior distribution of $\vec{\tau}|\mb h'$ can be heuristically approximated by a Gaussian distribution centered at $\bd \tau^{\text{MAP}}$. 

For the posterior distribution in $(\bd \tau, \bd \beta)$-variable, we have
\begin{align*}
p(\bd \tau, \bd \beta|\mb h') &\propto \exp\big(-\frac{1}{\sigma^2}\|\mb h'-\mb A(\bd \tau)\bd \beta\|_2^2-\bd \beta^\H\mb E^{-1}\bd \beta\big) \\
&=\exp\big(-(\bd \beta-\bd \beta^*)^\H\bd \Sigma^{-1}(\bd \beta-\bd \beta^*)\\
&\qquad\ \ \quad-\frac{1}{\sigma^2}\|\mb h'\|_2^2+(\bd\beta^*)^\H\bd \Sigma^{-1}\bd\beta^*\big),
\end{align*}
where
\[\bd \Sigma(\bd \tau)= (\mb A^\H\mb A/\sigma^2+\mb E^{-1})^{-1},\ \ \bd \beta^*(\bd \tau)=\bd \Sigma \mb A^\H \mb h'/\sigma^2. \] 
The marginal posterior for $\bd \tau$ is
\begin{align*}
&p(\bd \tau|\mb h')\propto \operatorname{det}(\mb \Sigma) \exp\big((\bd\beta^*)^\H\bd \Sigma^{-1}\bd\beta^*\big).
\end{align*}
Its MAP 
\[\bd \tau^{\text{MAP}}=\mathop{\arg\max}_{\bd\tau}\: \log\det(\mb \Sigma)+(\bd\beta^*)^\H\bd \Sigma^{-1}\bd\beta^*. \]
For the joint posterior $p(\bd \tau,\bd \beta|\mb h')$, the MAP is
\[\tilde{\bd \tau}^{\text{MAP}}=\mathop{\arg\max}_{\bd\tau} \: (\bd\beta^*)^\H\bd \Sigma^{-1}\bd\beta^*,\ \tilde{\bd \beta}^{\text{MAP}}=\bd \beta^*(\tilde{\bd \tau}^{\text{MAP}}).\]

When $\sigma\ll 1$ (i.e., the high SNR regime), $\bd\Sigma\sim O(\sigma^2)$, $\bd\beta^*\sim O(1)$, thus the term $\log\det(\bd\Sigma)$ is negligible compared with $(\bd\beta^*)^\H\bd \Sigma^{-1}\bd\beta^*$.  So we approximately have $\bd \tau^{\text{MAP}}\approx\tilde{\bd \tau}^{\text{MAP}}$, and the Taylor expansion of $\log p(\bd \tau|\mb h')$ around $\bd \tau^{\text{MAP}}$ gives
\begin{align*}
&\log p(\bd \tau|\mb h')\approx \log(p(\bd \tau^{\text{MAP}}))-\frac12(\bd\tau-\bd\tau^{\text{MAP}})^\H H(\bd\tau-\bd \tau^{\text{MAP}}),
\end{align*}
which means $p(\bd\tau|\mb h')\sim N(\bd \tau|\bd\tau^{\text{MAP}},H^{-1})$. 
We get a Gaussian approximation of the posterior with mean $\bd \tau^{\text{MAP}}$ and covariance $H$, where $H$ is the Hessian of $-\log p(\bd \tau|\mb h')$, or $-\log p(\bd \tau,\bd \beta^*(\bd \tau)|\mb h')$  equivalently. As a by-product, the BFGS iteration for the reduced loss $-\log p(\bd \tau,\bd \beta^*(\bd \tau)|\mb h')$ gives an approximation of inverse Hessian automatically.

We extrapolate the above approximations to general SNR cases from the consideration of practical algorithm design.

\section{Optimal Low-Rank Basis Under Uniform Prior}\label{sec:app1}

We will show that when $L=N_p=1$ and $\tau|\mb h'\sim \text{Uniform}[-\Delta \tau /2,\Delta \tau/2]$, $\bd \beta \sim CN(0,E)$, The optimal low-rank basis is DPSS in \cite{DPSS}.

Under the uniform prior of $\tau |\mb h'$, the covariance $\operatorname{Cov}(\mb A_0 \bd \beta)=E\cdot \text{Toeplitz}(\vec{u})$, where $\vec{u}$ is a vector with components $u_k=\sin (k\Delta f\Delta \tau)$, $k= 0:m-1$, $\text{Toeplitz}(\vec{u})$ is the Toeplitz matrix generated by vector $\vec{u}$. Denoting $c=m\Delta f \Delta \tau$, $\mb T(c):=c/m\cdot\text{Toeplitz}(u)$, we have the following properties:

\begin{enumerate}
\item $\mb T(c)$ is a positive definite matrix, whose eigenvalues and eigenvectors are $\lambda_k(c),\bd \phi_k(c)$, $k=0:m-1$. These eigenvectors are called DPSSs in \cite{DPSS}.
\item If $c<m$, $\lambda_k(c)\in (0,1)$, $\sum_k \lambda_k(c)=c$.
\item The top $c+o(c)$ eigenvalues are close to 1 and the remaining  eigenvalues decrease to 0 exponentially in $k$.
\item $c\to 0,\lambda_0(c)\to 1,\bd \phi_0(c)\to \mb 1$.
\item Replacing the posterior with $\text{Uniform}[\tau_0-\Delta \tau /2,\tau_0+\Delta \tau/2]$, the eigenvectors will become $\bd \phi_k(c)\odot \{e^{-\j 2\pi k\Delta f\tau_0}\}$ and the eigenvalues remain unchanged.
\end{enumerate}

One can calculate the channel estimation NMSE of the first $k$-bases LMMSE:
\begin{equation}\label{eq:nmse}
1-\sum_{i=0}^{k-1} \frac{(\lambda_i(c)/c)^2}{\lambda_i(c)/c+\sigma^2/mE}.
\end{equation}
The NMSE increases monotonically with respect to $c$. The parameter $c$ and the corresponding eigenvalues $\lambda_i(c) $ characterizes the error. It can be found that if $\lambda_i(c)/c\ll \sigma^2/mE$, the $i$-th basis contributes little to reduce the error. Hence we only need few top eigenvectors to obtain a sufficiently good estimate.

The above derivations can be generalized to multipath case as well. However, there will be no explicit form of the optimal low rank basis.

\section*{Acknowledgment}

The authors would like to thank Prof. An Liu for stimulating discussions.  They also acknowledge the support from National Key R\&D Program of China under grant 2021YFA1003301, and National Science Foundation of China under grants 12288101 and 12101230.

\bibliographystyle{IEEEtran}
\bibliography{ref}

\end{document}